\documentstyle[emulateapj,apjfonts]{article}

\def\Msun{\ifmmode M_{\odot} \else $M_{\odot}$\fi}
\def\Lsun{\ifmmode L_{\odot} \else $L_{\odot}$\fi}
\def\qo{\ifmmode q_{0} \else $q_{0}$\fi}
\def\Ho{\ifmmode H_{0} \else $H_{0}$\fi}

\lefthead{KRISS ET AL.}
\righthead{UV SPECTRUM OF NGC 7469}

\begin{document}
\submitted{To appear in the Astrophysical Journal}

\title{A High Signal-to-Noise UV Spectrum of NGC~7469:
New Support for Reprocessing of Continuum Radiation$^1$
}

\author{Gerard A. Kriss\altaffilmark{2,3},
Bradley M. Peterson\altaffilmark{4},
D. Michael Crenshaw\altaffilmark{5},
and
Wei Zheng\altaffilmark{3}
}
\altaffiltext{1}{
Based on observations with the NASA/ESA {\it Hubble Space Telescope},
obtained at the Space Telescope Science Institute,
which is operated by the Association of Universities for Research in Astronomy,
Inc., under NASA contract NAS5-26555.
These observations are associated with proposal ID 6747.
}
\altaffiltext{2}{Space Telescope Science Institute,
3700 San Martin Drive, Baltimore, MD 21218; gak@stsci.edu}
\altaffiltext{3}{Center for Astrophysical Sciences, Department of Physics and
Astronomy, The Johns Hopkins University, Baltimore, MD 21218--2686;
zheng@pha.jhu.edu}
\altaffiltext{4}{Department of Astronomy, The Ohio State University,
140 West 18th Avenue, Columbus, OH 43210; peterson@astronomy.ohio-state.edu}
\altaffiltext{5}{Catholic University of America and Laboratory for Astronomy and
Solar Physics, NASA Goddard Space Flight Center, Code 681, Greenbelt, MD 20771;
crenshaw@buckeye.gsfc.nasa.gov}

\setcounter{footnote}{0}

\begin{abstract}

From 1996 June 10 to 1996 July 29 the International AGN Watch monitored
the Seyfert 1 galaxy NGC~7469 using the International Ultraviolet Explorer,
the Rossi X-ray Timing Explorer and a network of ground-based observatories.
On 1996 June 18, in the midst of this intensive monitoring period, we
obtained a high signal-to-noise snapshot of the UV spectrum from
1150--3300 \AA\ using the Faint Object Spectrograph on the Hubble Space
Telescope.  This spectrum allows us to disentangle the UV continuum more
accurately from the broad wings of the emission lines,
to identify clean continuum windows free of contaminating emission
and absorption, and to deblend line complexes such as Ly$\alpha +${\sc N~v},
{\sc C~iv}$+$He~{\sc ii}$+${\sc O~iii}], Si~{\sc iii}]$+${\sc C~iii}],
and Mg~{\sc ii}$+$Fe~{\sc ii}.
Using the FOS spectrum as a template, we have fit and extracted line and
continuum fluxes from the IUE monitoring data.
The cleaner continuum extractions confirm the discovery of time delays
between the different UV continuum bands by Wanders et al.
Our new measurements show delays increasing with wavelength
for continuum bands centered at 1485 \AA, 1740 \AA\ and 1825 \AA\ relative
to 1315 \AA\ with delays of 0.09, 0.28 and 0.36 days, respectively.
Like many other Seyfert 1 galaxies, the UV spectrum of NGC~7469 shows
intrinsic, blue-shifted absorption in Ly$\alpha$, {\sc N~v} and {\sc C~iv}.
Soft X-ray absorption is also visible in archival ASCA X-ray spectra.
The strength of the UV absorption, however, is not compatible with a
single-zone model in which the same material absorbs both the UV and X-ray
light.  Similar to other Seyfert galaxies such as NGC~3516, the UV-absorbing
gas in NGC~7469 has a lower ionization parameter and column density than
the X-ray absorbing material.
While the UV and X-ray absorption does not arise in the same material,
the frequent occurrence of both associated UV absorption and X-ray warm
absorbers in the same galaxies suggests that the gas supply for each
has a common origin.

\end{abstract}


\section{Introduction}

For the last year of operations of the International Ultraviolet Explorer (IUE) 
the International AGN Watch successfully carried out
an intensive continuous monitoring campaign on the bright
Seyfert 1 galaxy NGC~7469 (\cite{Wanders97}).
During the course of these observations
we obtained a single high signal-to-noise UV spectrum covering 1150--3300 \AA\ 
using the Faint Object Spectrograph (FOS) on the Hubble Space Telescope (HST).
Simultaneous high-energy X-ray observations were obtained using the
Rossi X-ray Timing Explorer (RXTE) (\cite{Nandra98},
and a network of ground-based facilities obtained optical spectra
(\cite{Collier98}).
These data sets have been used to study the structure of the continuum
and line-emitting regions in NGC~7469 using reverberation mapping.

Previous IUE and ground-based campaigns that have applied reverberation
mapping techniques to the study of AGN broad-line regions (BLR) have
greatly illuminated our understanding of their structure
(see the review by \cite{Peterson93}).
The reverberation mapping method (\cite{Blandford82}) uses the
light-travel-time delayed response of the emission line clouds to
variations in the continuum to unravel the spatial and kinematic
structure of the BLR.
Campaigns on NGC~5548 and NGC~3783 using IUE
(\cite{Clavel91}; \cite{Reichert94})
and again on NGC~5548 using IUE and HST (\cite{Korista95}) have
determined that the BLR is smaller than single-zone photoionization models
have suggested, and that it is highly stratified.
Inner and outer radii differ by an order of magnitude, and higher ionization
lines are characteristically formed in the innermost regions.

\begin{deluxetable}{ c c c c c }
\tablecolumns{5}
\tablewidth{325pt}
\tablecaption{FOS Observations of NGC 7469 \label{tbl-obs}}
\tablehead{
\colhead{Root File Name} &
\colhead{Grating} &
\colhead{Start Time\tablenotemark{a}} &
\colhead{JD} &
\colhead{Integration Time} \\
\colhead{} &
\colhead{} &
\colhead{(UT)} &
\colhead{$(-2450000)$} &
\colhead{(s)}
}
\startdata
y3b60106t   &   G130H & 19:36:55 &  252.317  & 1200 \nl
y3b60107t   &   G130H & 20:54:11 &  252.371  & 2240 \nl
y3b6010at   &   G190H & 22:30:41 &  252.438  & 1650 \nl
y3b6010bt   &   G270H & 23:06:26 &  252.463  & \phn 300 \nl
\enddata
\tablenotetext{a}{
All observations occurred on 1996 June 18.
}
\end{deluxetable}

Analysis of the IUE data for the NGC~7469 campaign (\cite{Wanders97})
have led to similar results for its broad emission lines.
The most remarkable result, however, is the apparent detection of a time delay
in the response of different UV continuum windows.
The fluxes in bands centered at 1485 \AA, 1740 \AA, and 1825 \AA\ have
cross-correlation centroids with time delays of 0.21, 0.35, and
0.28 days with respect to the flux at 1315 \AA.
Monte Carlo simulations indicate probable errors of $\sim 0.07$ days
in measuring the delays.
Even longer delays ($\sim$1 day) are found for the optical continuum relative
to the UV (\cite{Collier98}).
A variety of explanations may lead to the observed effects.
The most interesting in terms of the overall structure of AGN is that the delays
are due to a continuum reprocessing zone near the central continuum source.
A more mundane possibility is that the delay is the result of contamination
of the flux in the continuum bands by a very broad emission feature such
as blended Fe~{\sc ii} emission or Balmer continuum emission.
While this can explain some portion of the UV-continuum delays, as we show
later, it is difficult to ascribe the lag of the optical continuum to
emission-line contamination.

The higher spectral resolution and higher S/N of the FOS spectrum of NGC~7469
allows a better assessment of the possible contaminants in the chosen
IUE continuum intervals.
By using the FOS spectrum as a template, a model of the line and continuum
emission features can be fitted to the series of IUE spectra.
Similar techniques were successfully used on the earlier NGC~3783
(\cite{Reichert94}) and NGC~5548 (\cite{Korista95}) campaigns.
This paper describes the FOS data for NGC~7469 and presents new line
and continuum flux measurements extracted from the IUE data.
In \S2 we present the FOS observations and the analysis of that spectrum.
In \S3 we describe how the template based on the FOS data was fit to the
time series of IUE spectra and present a new analysis of the line and
continuum variability based on these measurements.
In \S4 we discuss the UV and X-ray absorbing material in NGC~7469.
We discuss our results in \S5, and give a summary of our conclusions in \S6.

\section{FOS Observations}

We observed NGC~7469 on 1996 June 18 (UT) using gratings G130H, G190H,
and G270H on the blue side of the FOS.
These three spectra cover the wavelength range 1150--3300 \AA\ with a resolution
of $220~\rm km~s^{-1}$.
The start times and integration times of the observations are given in
Table 1.
To ensure high S/N, good photometry and accurate flat-fielding, we observed
through the 0.86$''$ aperture and acquired the target
using a precision peak-up sequence.
The last 5$\times$5 peak up was done using the 0.26$''$ sec aperture on
0.052$''$ centers.
Centering in the $0.86''$ circular aperture was better than 0.04$''$.
The peak flux seen through the 0.26$''$ aperture at the last peak-up position
has a ratio to that seen through the 0.86$''$ aperture used for the observation
consistent with that of a point source.
This should alleviate any concern that the spectrum might
be contaminated by starlight from a nuclear starburst region.

The standard pipeline calibration applied to the data gives good results.
The two G130H observations agree to within 0.5\% with each other.
A weighted average of these two spectra was taken to produce a mean
G130H spectrum.
The overlap regions between the G130H, G190H, and G270H spectra agree to better
than 1\%.  In the 14 separate groups read out for the G130H observation,
there is a variation of 3.8\% peak to peak.  It is smooth and non-random,
but could be an instrumental artifact such as thermal variations around the
orbit.
No renormalizations of the flux scale were applied to any of the spectra.

We used the low ionization Galactic absorption lines to correct the
wavelength scale of each observation, assuming that these features are
at zero velocity.  G130H required a 0.3 \AA\ shift to the
blue, as did G270H.  The G190H spectrum required no adjustment, but only
the Al~{\sc ii} $\lambda$1670 line is strong enough to measure reliably.
We estimate our wavelengths are accurate to $\sim$50 $\rm km~s^{-1}$.

The merged, flux-calibrated spectrum from the four separate observations
is shown in Fig.~1.
The S/N per pixel (0.25--0.50 \AA) is greater than 10 at all wavelengths
longward of 1200 \AA; per 1 \AA\, it exceeds 20.
In addition to the usual broad emission lines and blue continuum, note
the pronounced dip in the spectrum at 2200 \AA\ indicative of Galactic
extinction, and the broad blends of Fe~{\sc ii} emission that become apparent
longward of 2000 \AA.
NGC~7469 also shows high-ionization, intrinsic absorption lines.
These are shown in the {\sc C~iv} region in Fig.~2,
and they are also present in {\sc N~v} and in Ly$\alpha$.

\begin{figure*}[t]
\plotfiddle{"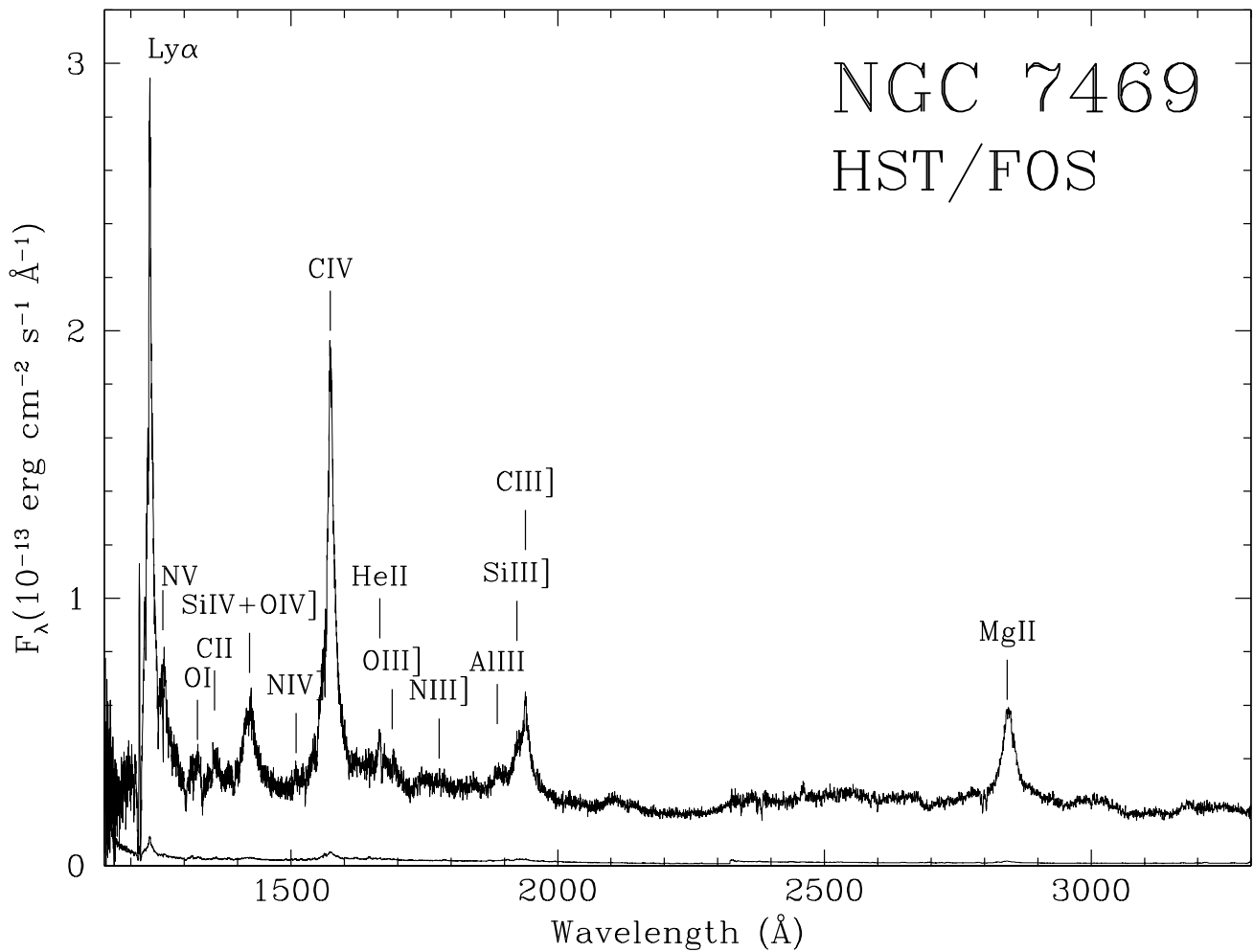"}{5.10 in} {-90}{70}{70}{-275}{410}
\caption{
Merged, flux-calibrated FOS spectrum of the Seyfert 1 galaxy NGC~7469.
The 1-$\sigma$ statistical errors are the thin line under the data.
The most prominent emission lines are labeled.
\label{fig1.ps}}
\end{figure*}

To model the lines and continuum in our spectrum, we
used the IRAF\footnote{
The Image Reduction and Analysis Facility (IRAF) is distributed by
the National Optical Astronomy Observatories, which is operated by the
Association of Universities for Research in Astronomy, Inc. (AURA) under
cooperative agreement with the National Science Foundation.
}
task {\tt specfit} (\cite{Kriss94})
to fit a model comprised of a reddened power law in $F_\lambda$

$$F_{\lambda} = F_{0} {\left( \lambda \over 1000 \right)}^{- \alpha},$$

\noindent
with extinction following the form given by
Cardelli, Clayton, \& Mathis (1989)\markcite{CCM89} with $\rm R_V = 3.1$,
multiple Gaussians for the emission lines,
single Gaussians for the Galactic and intrinsic absorption lines,
and a damped Lorentzian profile for the strong Galactic Ly$\alpha$ absorption.
We used the minimum number of Gaussian components necessary to
acceptably fit each emission line.  For the brightest emission lines,
three Gaussians (narrow, broad, and very broad) were typical.
An exception is the Si~{\sc iv}+{\sc O~iv}] $\lambda$1400 complex.
Here we allowed single Gaussians for each line in each multiplet set,
with all relative wavelengths linked in proportion to their vacuum values.
The widths of the two Si~{\sc iv} lines were linked to be identical, and
their flux ratio was fixed at 2:1.
The widths of the five {\sc O~iv}] lines were also linked to be identical,
but independent of the Si~{\sc iv} lines.
Their flux ratios were fixed in the proportion
0.1:0.2:1.0:0.4:0.1 as given by Osterbrock (1963)\markcite{Osterbrock63},
and the total {\sc O~iv}] flux varied independently of the Si~{\sc iv} flux.

Our fit covered the wavelength range 1170--3280 \AA, excluding an 8 \AA\ window
centered on geocoronal Ly$\alpha$ emission.
The best-fit $\chi^2$ is 7058 for 5479 points and 183 freely varying parameters.
We compute our error bars from the error matrix of the fit assuming a
$\Delta \chi^2 = 1$ for a single interesting parameter (\cite{Avni76}).

The best-fit continuum has a normalization of
$F_\lambda(1000\rm\AA)=
1.04 \pm 0.01 \times 10^{-13}~\rm erg~cm^{-2}~s^{-1}~\AA^{-1}$ and a
powerlaw index $\alpha = 0.977 \pm 0.003$.
The best fit extinction is $E(B - V) = 0.12 \pm 0.003$, and the column density
of the damped Ly$\alpha$ absorption is $\rm 3.5 \pm 0.2 \times 10^{20}~cm^{-2}$.
(Note that these errors are purely formal, statistical ones.  Systematic
errors due to our assumption of a continuum shape and due to the exclusion
of a large portion of the damped Ly$\alpha$ profile will be larger.)
Our measurements are in reasonable agreement with the properties of our own
Galaxy along the line of sight.
The Elvis, Lockman, \& Wilkes (1989)\markcite{Elvis89} {\sc H~i} survey of AGN
sight lines reports an {\sc H~i} column of
$4.82 \pm 0.17 \times 10^{20}~\rm cm^{-2}$.
Using a gas-to-dust ratio of
$N_{H I} / E(B - V) = 5.2 \times 10^{21}~\rm cm^{-2}$ (\cite{SVS85})
predicts $E(B - V) = 0.09$.

\vbox to 3.75in {
\vbox to 14pt{\vfill}
\plotfiddle{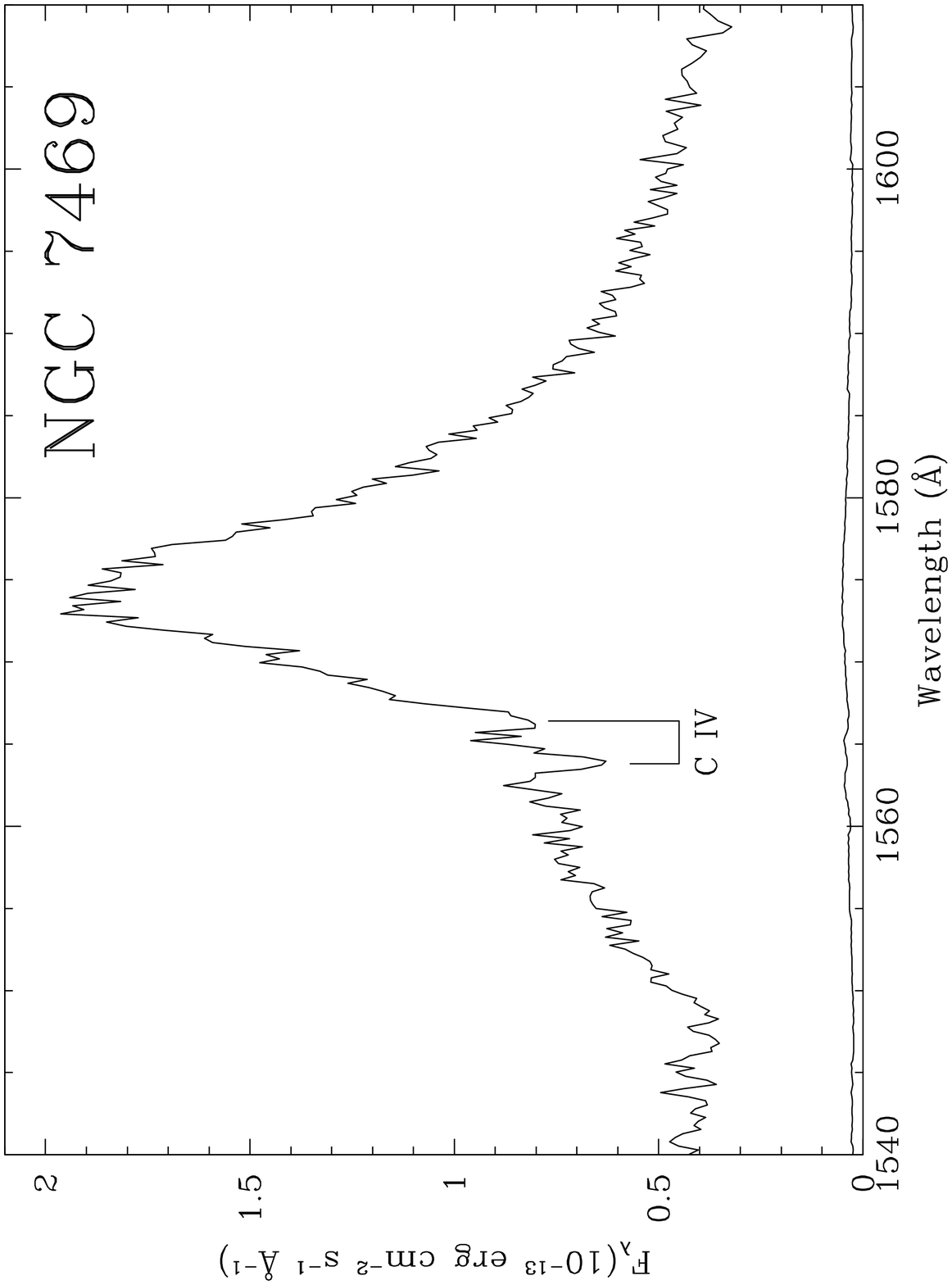}{2.6 in} {-90}{34}{34}{-127}{228}
\parbox{3.5in}{
\small\baselineskip 9pt
\footnotesize
\indent
{\sc Fig.}~2.---
The blueshifted intrinsic {\sc C~iv} absorption lines of NGC~7469
are visible in this plot of the {\sc C~iv} emission line region of the spectrum.
The 1-$\sigma$ statistical errors are the thin line under the data.
\label{fig2.ps}}
\vbox to 14pt{\vfill}
}
\setcounter{figure}{2}

The individual components for the emission lines are listed in
Table 2.
Parameters of the blueshifted intrinsic absorption features visible in
Ly$\alpha$, {\sc N~v}, and {\sc C~iv} are given in Table 3.
Galactic absorption features are listed in Table 4.
The tabulated line widths are not corrected for the instrumental resolution
of $220~\rm km~s^{-1}$.

\begin{center}
{\sc TABLE 2\\
Emission Line Fluxes in NGC 7469}
\vskip 4pt
\footnotesize
\begin{tabular}{ l c c c c }
\hline
\hline
Line & $\rm \lambda_{vac}$ & Flux$^{\rm a}$ & Velocity$^{\rm b}$ & FWHM \\
     & (\AA) &        & $\rm (km~s^{-1})$ & $\rm (km~s^{-1})$ \\
\hline
Ly$\alpha$ & 1215.67 & $258.0 \pm 16.8$ & \phantom{0}$-515 \pm \phantom{00}21$ & $\phantom{-0}967 \pm \phantom{00}44$ \\
Ly$\alpha$ & 1215.67 & $197.0 \pm 8.1$ & \phantom{0}$-268 \pm \phantom{00}33$ & $\phantom{-}2932 \pm \phantom{0}141$ \\
Ly$\alpha$ & 1215.67 & $242.0 \pm 19.3$ & \phantom{0}$-268 \pm \phantom{00}33$ & $10965 \pm \phantom{0}560$ \\
Ly$\alpha$~total & 1215.67 & $697.0 \pm 26.8$ & \nodata & \nodata \\
N~V        & 1240.15 & $10.4 \pm 1.5$ & $\phantom{-0}389 \pm \phantom{00}96$ & $\phantom{-}1598 \pm \phantom{00}63$ \\
N~V        & 1240.15 & $17.8 \pm 2.1$ & $\phantom{-0}389 \pm \phantom{00}96$ & $\phantom{-}4949 \pm \phantom{0}122$ \\
N~V        & 1240.15 & $48.2 \pm~5.7$ & $\phantom{-0}389 \pm \phantom{00}96$ & $12042 \pm \phantom{0}575$ \\
N~V~total  & 1240.15 & $76.4 \pm 2.5$ & \nodata & \nodata \\
Si~II      & 1260.45 & $\phantom{-}6.2 \pm 1.0$ & $\phantom{-0}939 \pm \phantom{0}132$ & $\phantom{-}2028 \pm \phantom{0}346$ \\
O~I        & 1304.35 & $21.0 \pm 1.6$ & \phantom{0}$-631 \pm \phantom{0}169$ & $\phantom{-}4618 \pm \phantom{0}373$ \\
C~II       & 1335.30 & $21.0 \pm 1.4$ & \phantom{0}$-165 \pm \phantom{0}116$ & $\phantom{-}3800 \pm \phantom{0}245$ \\
Si~IV      & 1393.76 & $45.2 \pm 4.5$ & \phantom{0}$-524 \pm \phantom{00}62$ & $11665 \pm \phantom{0}498$ \\
Si~IV      & 1402.77 & $22.6 \pm~2.3$ & \phantom{0}$-524 \pm \phantom{00}62$ & $11665 \pm \phantom{0}498$ \\
Si~IV~total & 1396.76 & $67.8 \pm 6.7$ & \nodata & \nodata \\
O~IV]~total & 1402.06 & $38.0 \pm 2.5$ & $\phantom{0}-524 \pm \phantom{00}62$ &  $\phantom{-}4002 \pm \phantom{0}300$ \\
N~IV]      & 1486.50 & $\phantom{-}2.9 \pm~0.5$ & \phantom{00}$-84 \pm \phantom{0}181$ & $\phantom{-}1420 \pm \phantom{0}307$ \\
C~IV       & 1549.05 & $66.2 \pm 4.5$ & $\phantom{-00}28 \pm \phantom{00}14$ & $\phantom{-}1598 \pm \phantom{00}63$ \\
C~IV       & 1549.05 & $166.0 \pm 2.0$ & \phantom{0}$-101 \pm \phantom{00}23$ & $\phantom{-}4949 \pm \phantom{0}122$ \\
C~IV       & 1549.05 & $160.0 \pm 6.2$ & \phantom{0}$-101 \pm \phantom{00}23$ & $12042 \pm \phantom{0}575$ \\
C~IV~total & 1549.05 & $392.2 \pm 7.9$ & \nodata & \nodata \\
Fe~II      & 1608.45 & $17.7 \pm 1.7$ & \phantom{0}$-391 \pm \phantom{0}190$ & $\phantom{-}5498 \pm \phantom{0}586$ \\
He~II      & 1640.50 & $\phantom{-}4.5 \pm 1.6$ & \phantom{00}$-81 \pm \phantom{0}200$ & $\phantom{-0}887 \pm \phantom{0}443$ \\
He~II      & 1640.50 & $18.5 \pm~0.7$ & \phantom{00}$-81 \pm \phantom{0}200$ & $\phantom{-}4949 \pm \phantom{0}122$ \\
He~II      & 1640.50 & $32.9 \pm~1.2$ & \phantom{00}$-81 \pm \phantom{0}200$ & $12042 \pm \phantom{0}575$ \\
He~II~total& 1640.50 & $55.8 \pm 1.8$ & \nodata & \nodata \\
O~III]     & 1663.48 & $\phantom{-}2.5 \pm~0.2$ & $\phantom{-0}391 \pm \phantom{00}97$ & $\phantom{-0}790 \pm \phantom{0}210$ \\
N~III]     & 1750.51 & $27.2 \pm 2.8$ & \phantom{0}$-210 \pm \phantom{0}399$ & $\phantom{-}8222 \pm \phantom{0}719$ \\
Al~III     & 1857.40 & $18.5 \pm 1.6$ & $\phantom{-0}412 \pm \phantom{0}157$ & $\phantom{-}4461 \pm \phantom{0}553$ \\
Si~III]    & 1892.03 & $19.1 \pm 3.9$ & \phantom{000}$-7 \pm \phantom{0}138$ & $\phantom{-}2470 \pm \phantom{0}160$ \\
C~III]     & 1908.73 & $\phantom{-}4.3 \pm~0.9$ & \phantom{00}$-77 \pm \phantom{00}29$ & $\phantom{-0}547 \pm \phantom{00}99$ \\
C~III]     & 1908.73 & $42.9 \pm 4.9$ & $\phantom{-0}142 \pm \phantom{0}136$ & $\phantom{-}3160 \pm \phantom{0}204$ \\
C~III]     & 1908.73 & $75.0 \pm 4.1$ & $\phantom{-0}142 \pm \phantom{0}136$ & $17050 \pm 1264$ \\
C~III]~total     & 1908.73 & $122.2 \pm 6.5$ & \nodata & \nodata \\
Mg~II      & 2798.74 & $\phantom{-}9.2 \pm 1.8$ & $\phantom{-00}56 \pm \phantom{00}15$ & $\phantom{0}1195 \pm \phantom{0}136$ \\
Mg~II      & 2798.74 & $90.0 \pm 1.8$ & $\phantom{-00}56 \pm \phantom{00}15$ & $\phantom{0}3426 \pm \phantom{00}72$ \\
Mg~II      & 2798.74 & $59.5 \pm 2.9$ & $\phantom{-00}56 \pm \phantom{00}15$ & $21393 \pm \phantom{0}662$ \\
Mg~II~total & 2798.74 & $158.7 \pm 3.9$ & \nodata & \nodata \\
\hline
\end{tabular}
\vskip 2pt
\parbox{3.5in}{
\small\baselineskip 9pt
\footnotesize
\indent
$\rm ^a$Observed flux in units of $\rm 10^{-14}~erg~cm^{-2}~s^{-1}$.\\
$\rm ^b$Velocity is relative to a systemic redshift of $cz = 4916~\rm km~s^{-1}$
(\cite{RC3}).
}
\end{center}
\setcounter{table}{2}

\begin{center}
{\sc TABLE 3\\
Intrinsic Absorption Lines in NGC 7469}
\vskip 4pt
\footnotesize
\begin{tabular}{ l c c c c }
\hline
\hline
Line & $\rm \lambda_{vac}$ & EW & Velocity & FWHM \\
     & (\AA)     & (\AA) & $\rm (km~s^{-1})$ & $\rm (km~s^{-1})$\\
\hline
Ly$\alpha$ & 1215.67 & $0.41 \pm~0.08$ & $-1870 \pm \phantom{00}17$ & $\phantom{-0}280 \pm \phantom{00}58$ \\
Ly$\alpha$ & 1215.67 & $5.04 \pm~0.30$ & \phantom{0}$-656 \pm \phantom{00}24$ & $\phantom{-}1439 \pm \phantom{00}61$ \\
N~V        & 1238.82 & $0.48 \pm~0.08$ & $-1834 \pm \phantom{00}20$ & $\phantom{-0}309 \pm \phantom{00}57$ \\
N~V        & 1242.80 & $0.24 \pm~0.07$ & $-1834 \pm \phantom{00}20$ & $\phantom{-0}309 \pm \phantom{00}57$ \\
C~IV       & 1548.19 & $0.45 \pm~0.05$ & $-1819 \pm \phantom{00}11$ & $\phantom{-0}275 \pm \phantom{00}27$ \\
C~IV       & 1550.77 & $0.35 \pm~0.04$ & $-1819 \pm \phantom{00}11$ & $\phantom{-0}275 \pm \phantom{00}27$ \\
\hline
\end{tabular}
\vskip 2pt
\parbox{3.5in}{
\small\baselineskip 9pt
\footnotesize
\indent
$\rm ^a$Velocity is relative to a systemic redshift of $cz = 4916~\rm km~s^{-1}$
(\cite{RC3}).
}
\end{center}
\setcounter{table}{3}

\section{IUE Spectra}

\subsection{Measuring Continuum and Emission-line Fluxes}

Analysis of the IUE spectra of NGC~7469 during the summer 1996 monitoring
campaign (\cite{Wanders97})
suggests a time delay in the responses of the longer wavelength
UV continuum bands relative to the shortest wavelength bin centered at
1315 \AA.  One possible explanation for these delays is that the IUE
measurements are not using pure continuum, and that light from broadly
distributed line emission in the spectrum might be contaminating the data.
Our fits to the FOS data allow us to examine the degree of contamination.
Fig.~3 shows the FOS spectrum of NGC~7469 scaled to provide a good
view of the continuum.  The bands used for the IUE measurements are
indicated, and the best-fit reddened powerlaw is also shown.
Note that only in the 1485 \AA\ band does the fitted continuum pass
through the actual data points.

\vbox to 3.8in {
\vbox to 14pt{\vfill}
\plotfiddle{"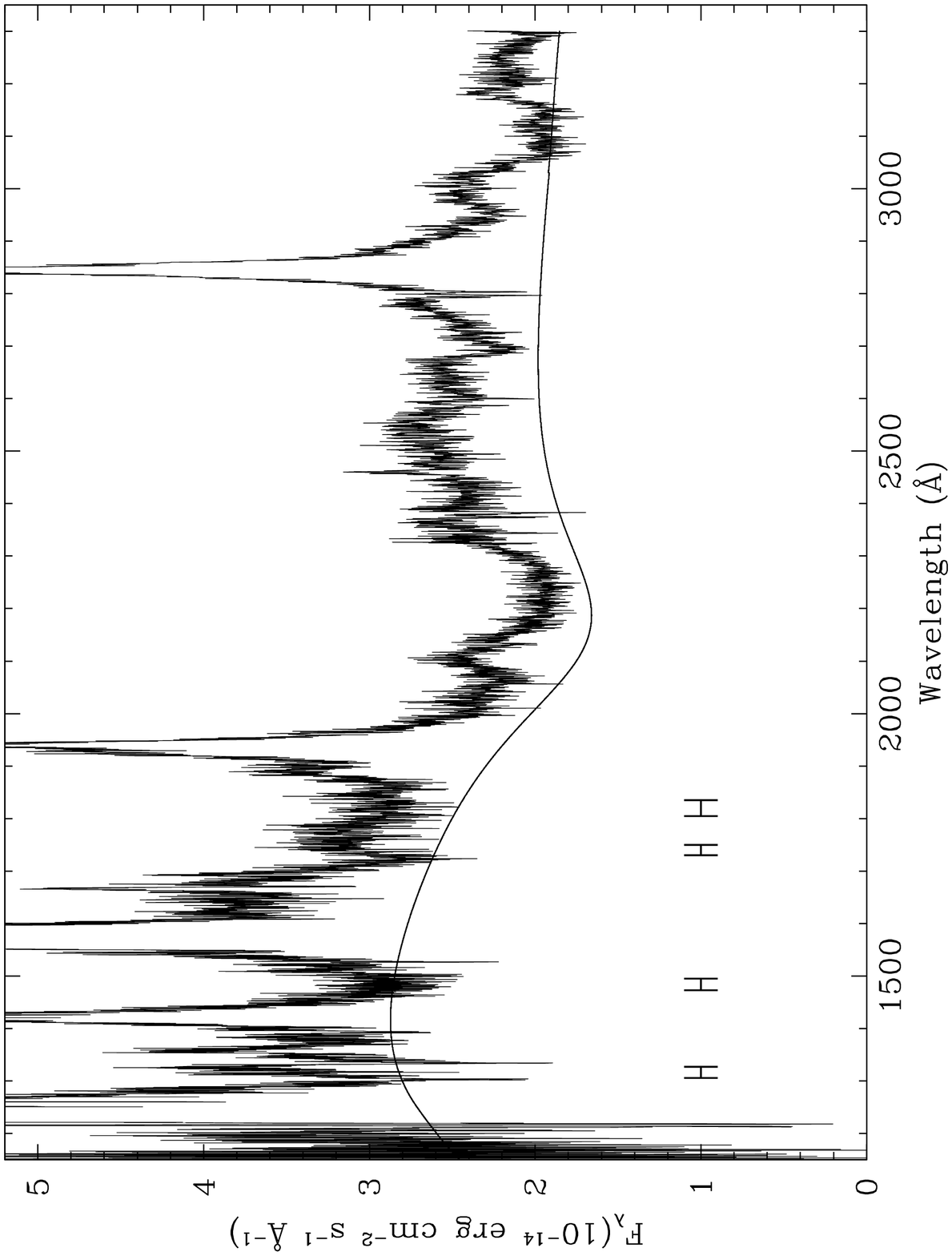"}{2.6 in} {-90}{34}{34}{-140}{200}
\parbox{3.5in}{
\small\baselineskip 9pt
\footnotesize
\indent
{\sc Fig.}~3.---
The best fit powerlaw continuum for the NGC 7469 is shown as the thin
solid line.  The dip at 2200 \AA\ and the downturn at the short wavelength
end reflect the extinction of $E(B - V) = 0.12$.
At most points in the spectrum, the blended wings of the broad emission lines
and Fe~{\sc ii} emission contribute a substantial amount of overlying flux.
The four ``continuum" windows used for measuring the fluxes in the IUE
spectra are shown as heavy solid bars.
\label{fig3.ps}}
\vbox to 14pt{\vfill}
}
\setcounter{figure}{3}

\begin{center}
{\sc TABLE 4\\
Galactic Absorption Lines in NGC 7469}
\footnotesize
\begin{tabular}{ l c c c c }
\hline
\hline
Line & $\rm \lambda_{vac}$ & EW & Velocity & FWHM \\
     & (\AA)     & (\AA) & $\rm (km~s^{-1})$ & $\rm (km~s^{-1})$\\
\hline
Si~II      & 1190.42 & $0.16 \pm~0.11$ & $\phantom{-00}20 \pm \phantom{00}54$ & $\phantom{-0}211 \pm \phantom{0}150$ \\
Si~II      & 1193.14 & $0.12 \pm~0.14$ & $\phantom{-00}80 \pm \phantom{00}93$ & $\phantom{-0}211 \pm \phantom{0}150$ \\
N~I        & 1200.16 & $0.39 \pm~0.17$ & \phantom{0}$-117 \pm \phantom{00}60$ & $\phantom{-0}303 \pm \phantom{0}119$ \\
Si~III     & 1206.50 & $1.07 \pm~0.22$ & \phantom{0}$-186 \pm \phantom{00}45$ & $\phantom{-0}593 \pm \phantom{0}112$ \\
S~II       & 1250.58 & $0.23 \pm~0.07$ & \phantom{0}$-141 \pm \phantom{000}0$ & $\phantom{-0}303 \pm \phantom{00}41$ \\
S~II       & 1253.00 & $0.21 \pm~0.06$ & $\phantom{-00}60 \pm \phantom{00}53$ & $\phantom{-0}303 \pm \phantom{00}41$ \\
S~II       & 1259.52 & $0.26 \pm~0.07$ & \phantom{00}$-26 \pm \phantom{00}50$ & $\phantom{-0}303 \pm \phantom{00}41$ \\
Si~II      & 1260.42 & $0.58 \pm~0.09$ & $\phantom{-000}7 \pm \phantom{00}20$ & $\phantom{-0}303 \pm \phantom{00}41$ \\
O~I        & 1302.17 & $0.41 \pm~0.07$ & \phantom{000}$-2 \pm \phantom{00}28$ & $\phantom{-0}303 \pm \phantom{00}41$ \\
Si~II      & 1304.37 & $0.36 \pm~0.07$ & \phantom{00}$-60 \pm \phantom{00}31$ & $\phantom{-0}303 \pm \phantom{00}41$ \\
C~II       & 1334.53 & $0.32 \pm~0.16$ & \phantom{0}$-290 \pm \phantom{00}47$ & $\phantom{-0}433 \pm \phantom{0}112$ \\
C~II       & 1335.69 & $0.68 \pm~0.19$ & \phantom{0}$-290 \pm \phantom{00}47$ & $\phantom{-0}433 \pm \phantom{0}112$ \\
Si~IV      & 1393.76 & $0.97 \pm~0.16$ & \phantom{00}$-97 \pm \phantom{00}58$ & $\phantom{-0}950 \pm \phantom{0}140$ \\
Si~IV      & 1402.77 & $0.62 \pm~0.15$ & \phantom{00}$-68 \pm \phantom{00}91$ & $\phantom{-0}950 \pm \phantom{0}140$ \\
Si~II      & 1527.17 & $0.45 \pm~0.06$ & \phantom{00}$-55 \pm \phantom{00}21$ & $\phantom{-0}314 \pm \phantom{00}91$ \\
C~IV       & 1548.19 & $0.38 \pm~0.07$ & \phantom{0}$-325 \pm \phantom{00}44$ & $\phantom{-0}390 \pm \phantom{00}57$ \\
C~IV       & 1550.77 & $0.43 \pm~0.07$ & \phantom{0}$-325 \pm \phantom{00}44$ & $\phantom{-0}390 \pm \phantom{00}57$ \\
Fe~II      & 1608.45 & $0.27 \pm~0.07$ & $\phantom{-00}62 \pm \phantom{00}45$ & $\phantom{-0}314 \pm \phantom{00}91$ \\
Al~II      & 1670.79 & $0.52 \pm~0.31$ & \phantom{00}$-57 \pm \phantom{00}17$ & $\phantom{-0}314 \pm \phantom{00}91$ \\
Fe~II      & 2344.21 & $0.32 \pm~0.10$ & \phantom{000}$-3 \pm \phantom{00}89$ & $\phantom{-0}518 \pm \phantom{00}58$ \\
Fe~II      & 2374.46 & $1.05 \pm~0.16$ & $\phantom{-00}20 \pm \phantom{00}39$ & $\phantom{-0}518 \pm \phantom{00}58$ \\
Fe~II      & 2382.77 & $1.11 \pm~0.18$ & \phantom{0}$-113 \pm \phantom{00}36$ & $\phantom{-0}518 \pm \phantom{00}58$ \\
Fe~II      & 2586.65 & $0.78 \pm~0.11$ & \phantom{0}$-104 \pm \phantom{00}44$ & $\phantom{-0}518 \pm \phantom{00}58$ \\
Fe~II      & 2600.17 & $0.60 \pm~0.10$ & \phantom{00}$-48 \pm \phantom{00}46$ & $\phantom{-0}518 \pm \phantom{00}58$ \\
Mg~II      & 2796.35 & $0.57 \pm~0.07$ & \phantom{000}$-1 \pm \phantom{00}11$ & $\phantom{-0}230 \pm \phantom{00}27$ \\
\hline
\end{tabular}
\end{center}
\setcounter{table}{4}

\begin{figure*}[t]
\plotfiddle{"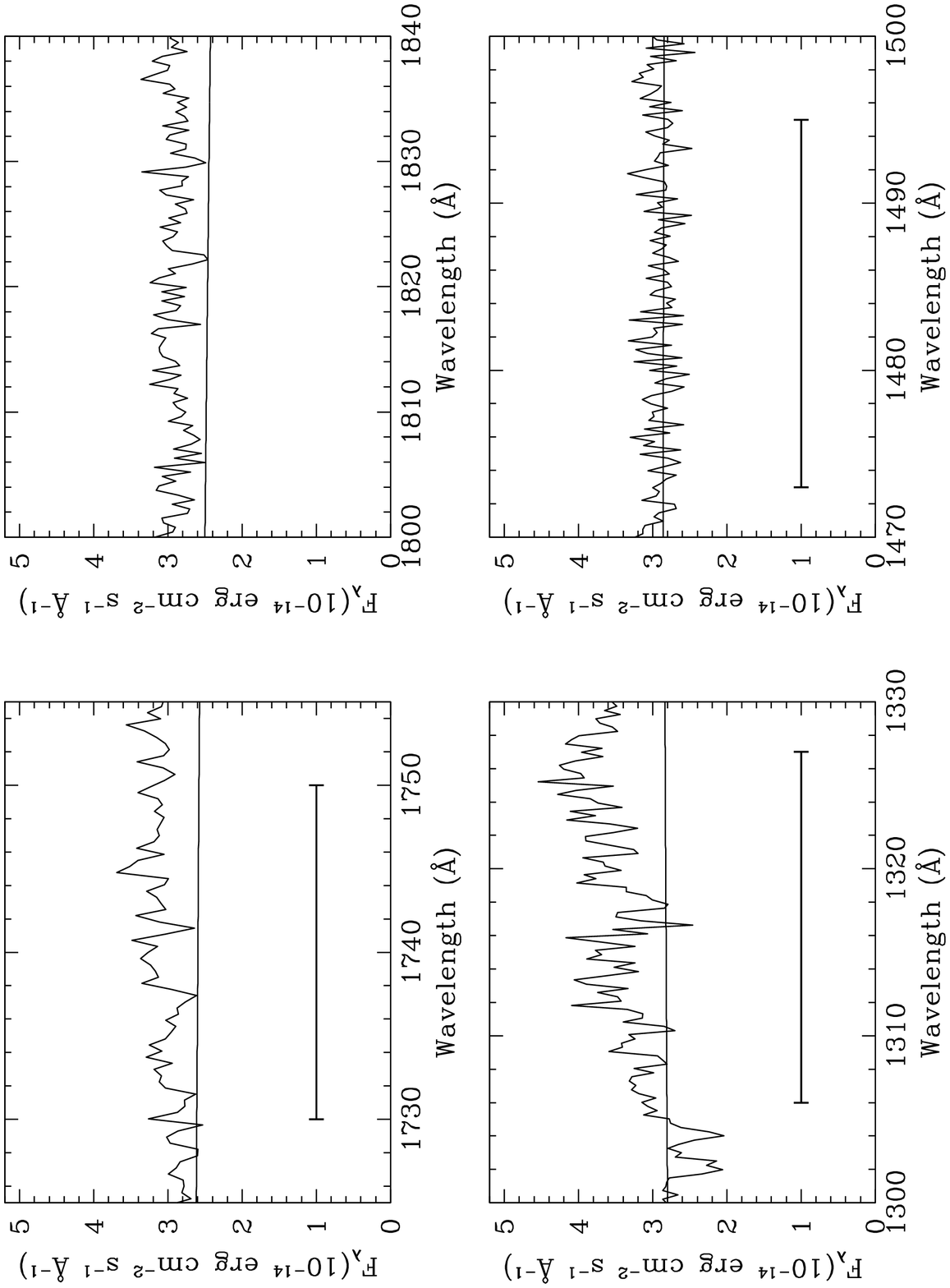"}{5.10 in} {-90}{70}{70}{-275}{410}
\caption{
The four panels give a close up view of the continuum windows used for the
IUE flux measurements.  The heavy solid bars show the wavelength intervals
used.  Upper left: $F_\lambda$(1740\AA); upper right: $F_\lambda$(1825\AA);
lower left: $F_\lambda$(1315\AA); lower right: $F_\lambda$(1485\AA).
{\sc O~i} $\lambda 1304$ emission contaminates the  $F_\lambda$(1315\AA) window,
and blended Fe~{\sc ii} emission contaminates the $F_\lambda$(1740\AA) and the
$F_\lambda$(1825\AA) windows.  Only the $F_\lambda$(1485\AA) is pure continuum.
\label{fig4.ps}}
\end{figure*}
\setcounter{figure}{4}

Fig.~4 illustrates the degree of contamination in more detail with
magnified views of each of the continuum bands along with the fitted
continuum.  The shortest wavelength bin centered on 1315 \AA\ is
contaminated by {\sc O~i} $\lambda$1304 emission.  Only 78\% of the flux
in the IUE measurement bin is from the fitted continuum.
The two longest wavelengths bins, 1740 \AA\ and 1825 \AA, are each
contaminated by Fe~{\sc ii} emission.  Within the IUE measurement bins
the percentage of flux due to the fitted continuum is 81\% and 85\%,
respectively.
The bin at 1485 \AA\ is relatively clean.
99\% of the flux in the IUE measurement bin is due to the fitted continuum.

Our new analysis of the IUE spectra from the 1996 campaign aims to obtain
clean measurements of the continuum and deblended measurements of the
emission lines using the FOS spectrum as a template.
For the fits described below we used only the TOMSIPS-extracted spectra
described by Wanders et al. (1997)\markcite{Wanders97}.  As noted by Wanders et
al., these spectra appear slightly smoother to the eye and have slightly smaller
error bars.  In the initial analysis by Wanders et al., both the TOMSIPS and
NEWSIPS data gave similar results.
Table 1 of Wanders et al. (1997)\markcite{Wanders97} logs 219 spectra
obtained using IUE.  We have restricted our analysis to the 207 spectra
unaffected by target centering problems or short exposure times, i.e.,
we have eliminated all spectra with notes 1--4 from Table 1 of Wanders et al.
The mean of these spectra is shown as Figure~1 of Wanders et al.

Using the model fit to the FOS data in \S2, we developed a template for
fitting the series of IUE spectra by first fitting the mean IUE spectrum.
The best-fit continuum has a normalization of
$F_\lambda(1000\rm\AA)=
1.37 \times 10^{-13}~\rm erg~cm^{-2}~s^{-1}~\AA^{-1}$ and a
powerlaw index $\alpha = 0.913 \pm 0.003$, close to the shape and intensity
of the FOS snapshot.
Emission-line fluxes, wavelengths, and widths are listed in
Table 5, and absorption-line parameters are given in
Table 6.
The resulting best-fit model is shown overlayed on the mean IUE spectrum
in Fig.~5; the residuals shown in the lower panel have an
rms of a few percent of the spectral intensity.

Due to the lower resolution and lower S/N of the individual IUE spectra,
numerous constraints were imposed on the use of this template for the fits
to the individual spectra.
For example, the wavelengths of weak emission lines were tied to
that of {\sc C~iv}$\lambda 1549$ by the ratios of their laboratory values;
the widths of weak lines were fixed at the values obtained in a fit to
the mean IUE spectrum; the wavelengths of multiple components
of strong lines were all fixed at the same wavelength; the parameters of
all absorption features were fixed at the values obtained in the fit to the
mean IUE spectrum. This left 44 free parameters for the fit to each spectrum:
the power-law normalization and exponent; the fluxes of the individual emission
lines; and the wavelengths and widths of the bright emission lines.
Each spectrum was then fit using {\tt specfit}.
To provide initial parameters for each fit, we used the best fit to the
mean spectrum as a starting point.  The continuum normalization was then scaled
by the ratio of the 1485 \AA\ continuum flux to the same continuum flux in the
mean spectrum; line fluxes were scaled by the ratio of the integrated net
{\sc C~iv} flux to the same flux measured in the mean spectrum;
line wavelengths were shifted by the location of the peak of the {\sc C~iv}
line relative to its location in the mean spectrum.

\begin{center}
{\sc TABLE 5\\
Emission Line Fluxes in the Mean\\
IUE Spectrum of NGC 7469}
\small
\begin{tabular}{ l c c c c }
\hline
\hline
Line & $\rm \lambda_{vac}$ & Flux$^{\rm a}$ & Velocity$^{\rm b}$ & FWHM\\
    & (\AA)    &                & $\rm ( km~s^{-1} )$ & $\rm ( km~s^{-1} )$ \\
\hline
Ly$\alpha$ & 1215.67 & \phantom{0}86.4 & $-$622 & \phantom{0}1122 \\
Ly$\alpha$ & 1215.67 & 156.0 & $-$622 & \phantom{0}2144 \\
Ly$\alpha$ & 1215.67 & 150.0 & $-$622 & \phantom{0}8134 \\
Ly$\alpha$ total & 1215.67 & 392.4 & \nodata & \nodata \\
N~V        & 1240.15 & \phantom{0}25.9 & $-$513 & \phantom{0}2405 \\
N~V        & 1240.15 & \phantom{0}22.7 & $-$513 & \phantom{0}6292 \\
N~V        & 1240.15 & \phantom{0}61.3 & $-$513 & 14764 \\
N~V total  & 1240.15 & 109.9 & \nodata & \nodata \\
Si~II      & 1260.45 & \phantom{00}3.0 & \phantom{0}747 & \phantom{0}3000 \\
O~I        & 1304.35 & \phantom{0}13.6 & $-$823 & \phantom{0}4700 \\
C~II       & 1335.30 & \phantom{0}13.8 & $-$322 & \phantom{0}3850 \\
Si~IV      & 1393.76 & \phantom{0}39.9 & $-$881 & 11521 \\
Si~IV      & 1402.77 & \phantom{0}20.0 & $-$871 & 11521 \\
Si~IV total& 1396.76 & \phantom{0}59.9 & \nodata & \nodata \\
O~IV]      & 1402.06 & \phantom{0}20.8 & $-$303 &  \phantom{0}3510 \\
N~IV]      & 1486.50 & \phantom{00}1.1 & $-$384 & \phantom{0}1000 \\
C~IV       & 1549.05 & \phantom{0}79.0 & $-$103 & \phantom{0}2405 \\
C~IV       & 1549.05 & 137.0 & $-$103 & \phantom{0}6292 \\
C~IV       & 1549.05 & 126.0 & $-$103 & 14764 \\
C~IV total & 1549.05 & 342.0 & \nodata & \nodata \\
He~II      & 1640.50 & \phantom{00}8.0 & $-$153 & \phantom{0}1716 \\
He~II      & 1640.50 & \phantom{0}22.0 & $-$153 & \phantom{0}6292 \\
He~II      & 1640.50 & \phantom{0}39.2 & $-$153 & 14764 \\
He~II total& 1640.50 & \phantom{0}69.2 & \nodata & \nodata \\
O~III]     & 1663.48 & \phantom{00}0.5 & \phantom{0}412 & \phantom{0}1795 \\
N~III]     & 1750.51 & \phantom{0}22.4 & $-$783 & \phantom{0}7821 \\
Al~III     & 1857.40 & \phantom{00}9.4 & $-$192 & \phantom{0}4387 \\
Si~III]    & 1892.03 & \phantom{0}14.4 & $-$635 & \phantom{0}2557 \\
C~III]     & 1908.73 & \phantom{00}3.0 & $-$460 & \phantom{0}1063 \\
C~III]     & 1908.73 & \phantom{0}35.2 & $-$460 & \phantom{0}3143 \\
C~III]     & 1908.73 & \phantom{0}57.7 & $-$460 & 17552 \\
C~III] total& 1908.73 & \phantom{0}95.9 & \nodata & \nodata \\
\hline
\end{tabular}
\vskip 2pt
\parbox{3.5in}{
\small\baselineskip 9pt
\footnotesize
\indent
$\rm ^a$Observed flux in units of $\rm 10^{-14}~erg~cm^{-2}~s^{-1}$.\\
$\rm ^b$ Velocity is relative to a systemic redshift of
$cz = 4916~\rm km~s^{-1}$ (\cite{RC3}).
}
\end{center}
\setcounter{table}{5}

Using the best-fit parameter values for each spectrum, we derived fluxes
for the quantities of interest.
Initial error bars were assigned based on the statistical 1-$\sigma$ values
obtained from {\tt specfit}. Final error bars were calculated using a
procedure common to our previous work in International AGN Watch campaigns.
We conservatively assume that there is no variation in flux between two
data points with a time separation $< 0.25$ d.
(The mean separation between observations is 0.23 d.)
We then scale the initial error bars so that
their mean fractional uncertainty is equal to the root-mean-square (rms)
of the distribution of flux ratios for all data pairs in the time series
with $\Delta t < 0.25$ d (\cite{RP97}; \cite{Wanders97}).
Note that the resulting error bars are an upper limit if there is any
residual intrinsic variability on timescales shorter than successive
observations in the time series.
The derived fluxes and errors are shown as light curves described
in the next section.
The actual data points and error bars can be obtained from the
International AGN Watch web site at the URL
{\tt http://www.astronomy.ohio-state.edu/$\sim$agnwatch/\#dat}.

\begin{center}
{\sc TABLE 6\\
Absorption Lines in the Mean\\
IUE Spectrum of NGC 7469}
\vskip 4pt
\small
\begin{tabular}{ l c c c c }
\hline
\hline
Line & $\rm \lambda_{vac}$ & EW    & Velocity            & FWHM \\
     & (\AA)               & (\AA) & $\rm ( km~s^{-1} )$ & $\rm ( km~s^{-1} )$\\
\hline
Ly$\alpha$ & 1215.67 & 0.39 & $-$1898\tablenotemark{a} & \phantom{0}1500 \\
Ly$\alpha$ & 1215.67 & 5.06 & \phantom{0}$-$661\tablenotemark{a} & \phantom{0}2000 \\
N~V        & 1238.82 & 1.17 & $-$2215\tablenotemark{a} & \phantom{0}1455 \\
N~V        & 1242.80 & 1.72 & $-$1595\tablenotemark{a} & \phantom{0}1455 \\
C~IV       & 1548.19 & 0.46 & $-$1851\tablenotemark{a} & \phantom{00}990 \\
C~IV       & 1550.77 & 0.36 & $-$1841\tablenotemark{a} & \phantom{00}990 \\
           &         &      &         &                 \\
S~II       & 1250.58 & 0.18 & \phantom{0}$-$141\tablenotemark{b} & \phantom{0}1455 \\
S~II       & 1253.00 & 0.20 & \phantom{000}36\tablenotemark{b} & \phantom{0}1455 \\
S~II       & 1259.52 & 0.25 & \phantom{00}$-$33\tablenotemark{b} & \phantom{0}1380 \\
Si~II      & 1260.42 & 0.56 & \phantom{0000}2\tablenotemark{b} & \phantom{0}1380 \\
O~I        & 1302.17 & 0.47 & \phantom{0}$-$840\tablenotemark{b} & \phantom{0}1255 \\
Si~II      & 1304.37 & 0.47 & \phantom{0}$-$899\tablenotemark{b} & \phantom{0}1255 \\
C~II       & 1334.53 & 0.25 & \phantom{0}$-$892\tablenotemark{b} & \phantom{0}1170 \\
C~II       & 1335.69 & 0.50 & \phantom{0}$-$884\tablenotemark{b} & \phantom{0}1170 \\
Si~IV      & 1393.76 & 0.51 & \phantom{0}$-$391\tablenotemark{b} & \phantom{0}1638 \\
Si~IV      & 1402.77 & 0.32 & \phantom{0}$-$391\tablenotemark{b} & \phantom{0}1638 \\
Si~II      & 1527.17 & 0.41 & \phantom{0}$-$585\tablenotemark{b} & \phantom{00}990 \\
C~IV       & 1548.19 & 0.34 & \phantom{0}$-$380\tablenotemark{b} & \phantom{00}990 \\
C~IV       & 1550.77 & 0.54 & \phantom{0}$-$369\tablenotemark{b} & \phantom{00}990 \\
Fe~II      & 1608.45 & 0.23 & \phantom{000}58\tablenotemark{b} & \phantom{00}995 \\
Al~II      & 1670.79 & 0.39 & \phantom{00}$-$20\tablenotemark{b} & \phantom{0}1000 \\
\hline
\end{tabular}
\vskip 2pt
\parbox{3.5in}{
\small\baselineskip 9pt
\footnotesize
\indent
$\rm ^a$Intrinsic absorption feature.
Velocity is relative to a systemic redshift of $cz = 4916~\rm km~s^{-1}$
(\cite{RC3}).\\
$\rm ^b$Galactic feature.  Velocity is heliocentric.
}
\end{center}
\setcounter{table}{6}

\begin{figure*}[t]
\plotfiddle{"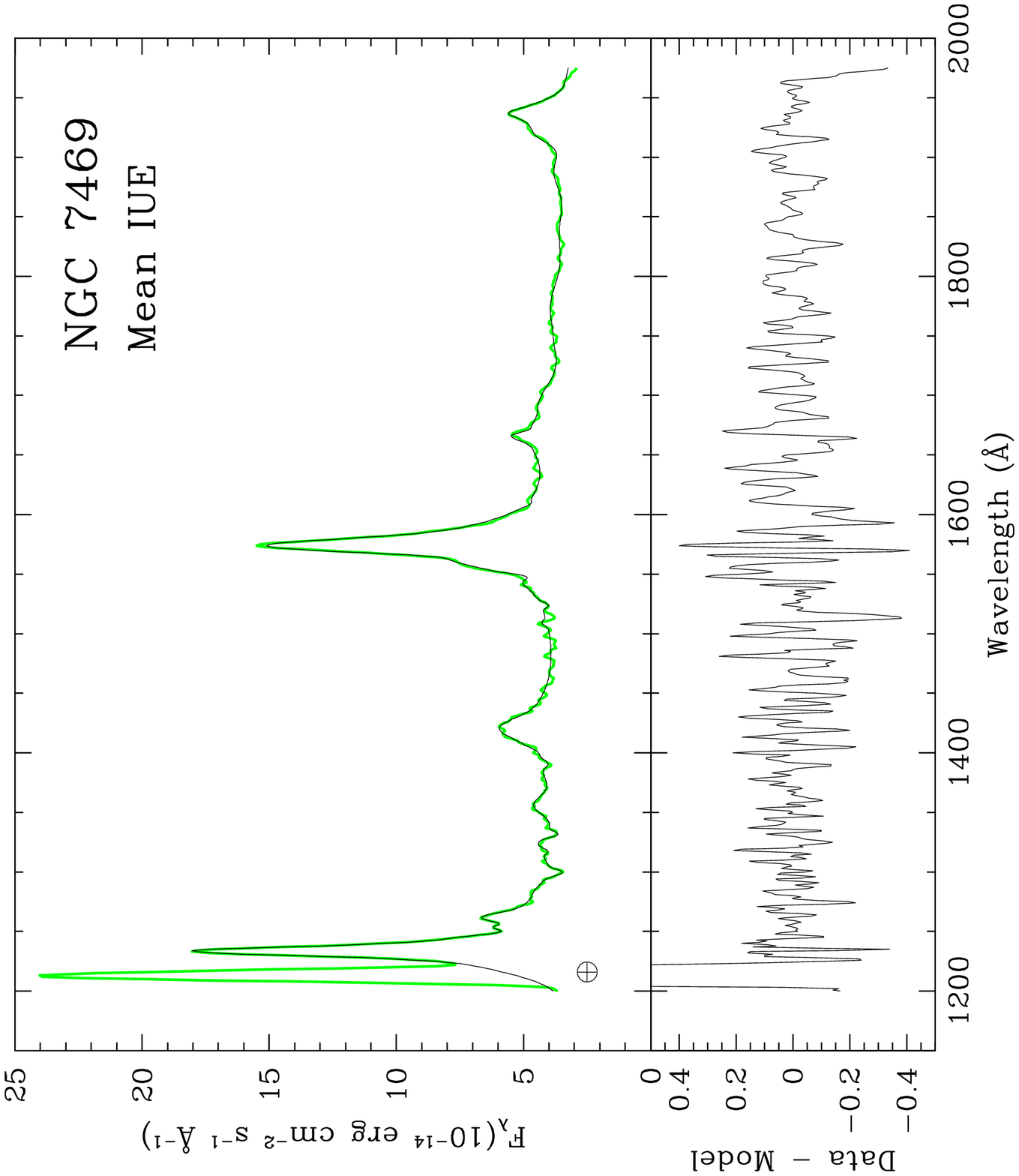"}{4.70 in} {-90}{61}{61}{-250}{370}
\caption{
Mean IUE spectrum of the Seyfert 1 galaxy NGC~7469 (gray curve) overlayed
with the best-fit model based on the FOS spectrum (thin black curve).
Geocoronal Ly$\alpha$ emission is indicated by $\earth$.
The lower panel shows the residuals to the fit.
\label{fig5.ps}}
\end{figure*}

We note that our use of a global power-law model for the underlying continuum
means that not all the continuum flux measurements we tabulate are statistically
independent.
The power-law model contains only two free parameters, its normalization and
the spectral index.  Thus in effect only two of the continuum fluxes suffice
to describe the data set at a single point in time.

\subsection{Continuum and Emission-line Light Curves}

The newly derived continuum light curves are shown in Fig.~6.
These are quite similar to the original data presented in Wanders et al.
All curves show the 10--15 day ``events" superposed on a gradual
decrease in flux from the start to the end of the campaign.
There are subtle differences, however, that are only apparent in a ratio between
the new measurements and the originals.
Light curves of these ratios are shown in Fig.~7.
All four light curves show slight differences from the originals
throughout the ``event" centered on day 280.  The most apparent differences
are in the light curve for F(1825 \AA), which shows departures from the original
surrounding all peaks in the light curve. The sense of the difference
is that when the source is brighter, more of the 1825 \AA\ flux
is due to continuum light.

The emission-line light curves are also quite similar
to those of Wanders et al.
These are shown in Figures~8--10.
Note that our deblending process has recovered more signal in the {\sc N~v}
and the He~{\sc ii} light curves.  None of the weaker lines, however,
show any strong correlation with the events in the continuum light curves.

\begin{figure*}[t]
\plotfiddle{"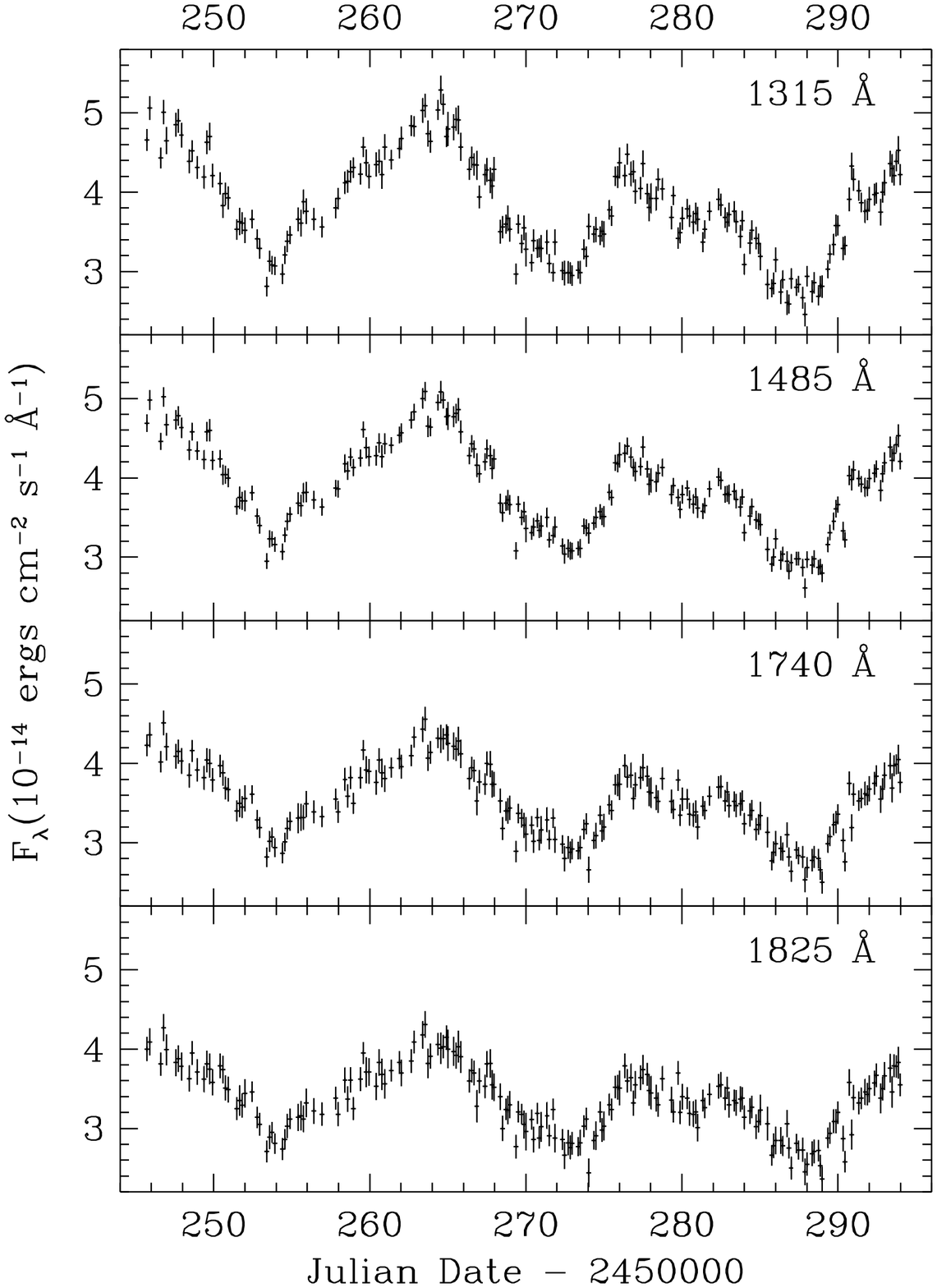"}{5.80 in} {0}{63}{63}{-210}{-47}
\caption{
Light curves for the continuum fluxes from the 1996 campaign on NGC~7469
extracted from the IUE spectra using the FOS spectrum as a template.
\label{fig6.ps}
}
\end{figure*}

\begin{figure*}[t]
\plotfiddle{"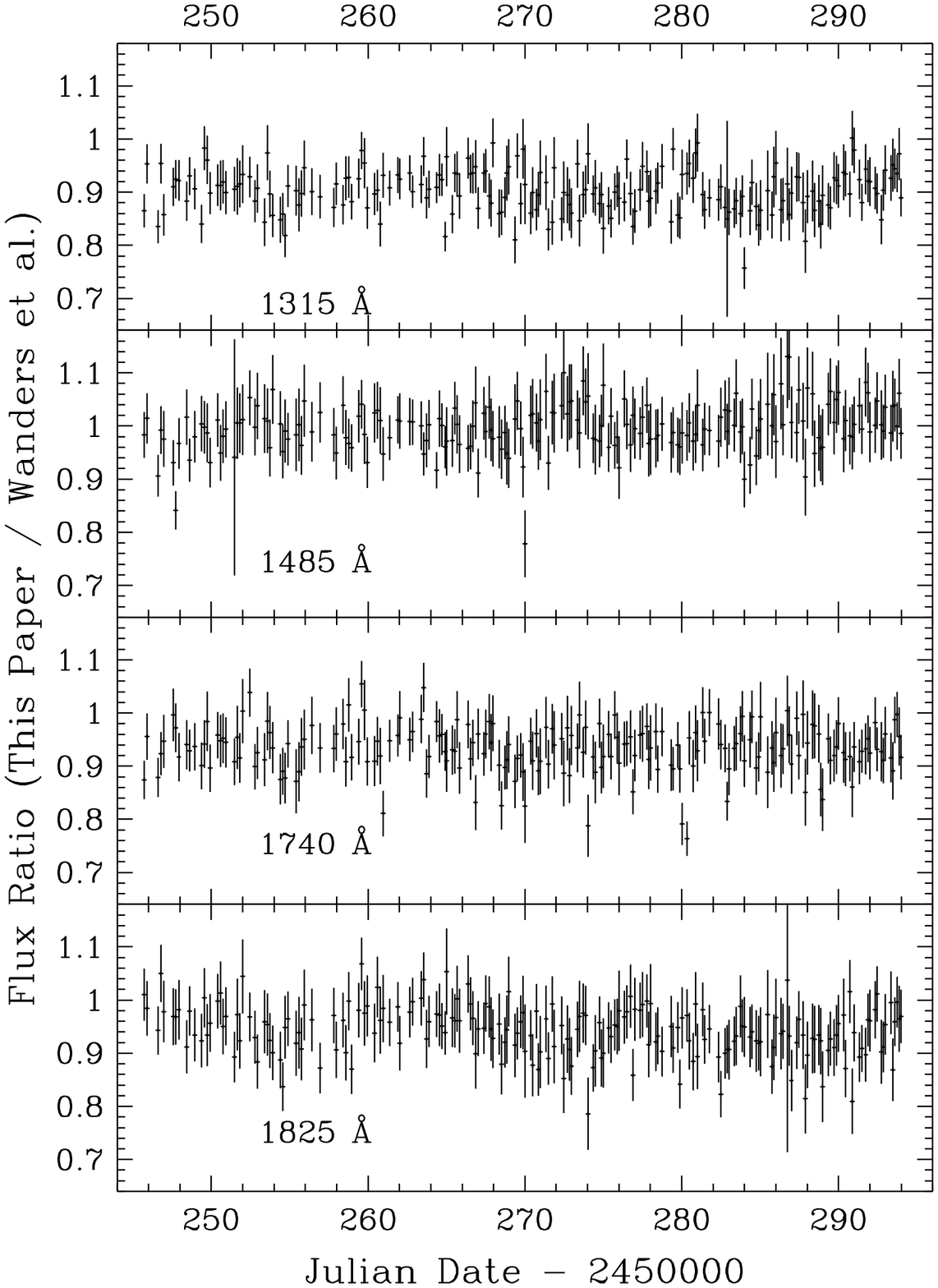"}{6.40 in} {0}{70}{70}{-210}{-47}
\caption{
Light curves of the ratios of the newly measured continuum fluxes shown in
Figure 6 to the original continuum fluxes from Wanders et al. (1997).
\label{fig7.ps}
}
\end{figure*}

\begin{figure*}[t]
\plotfiddle{"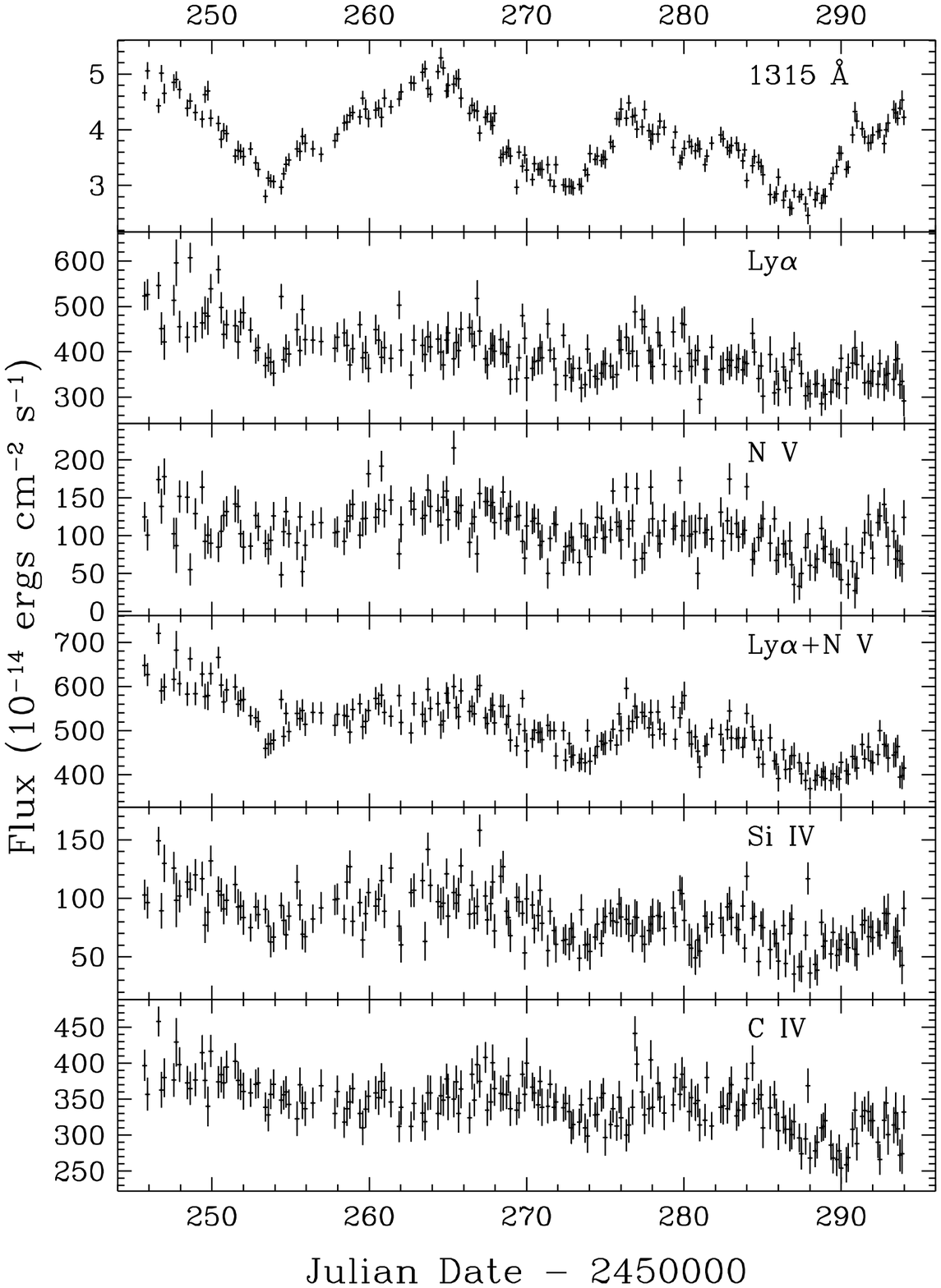"}{6.40 in} {0}{70}{70}{-210}{-47}
\caption{
Light curves for the strongest emission lines from the 1996 campaign on NGC~7469
extracted from the IUE spectra using the FOS spectrum as a template.
For comparison, the top panel gives the F(1315\AA) continuum light curve
(in units of $10^{-14}~\rm ergs~cm^{-2}~s^{-1}~\AA^{-1}$).
\label{fig8.ps}
}
\end{figure*}

\begin{figure*}[t]
\plotfiddle{"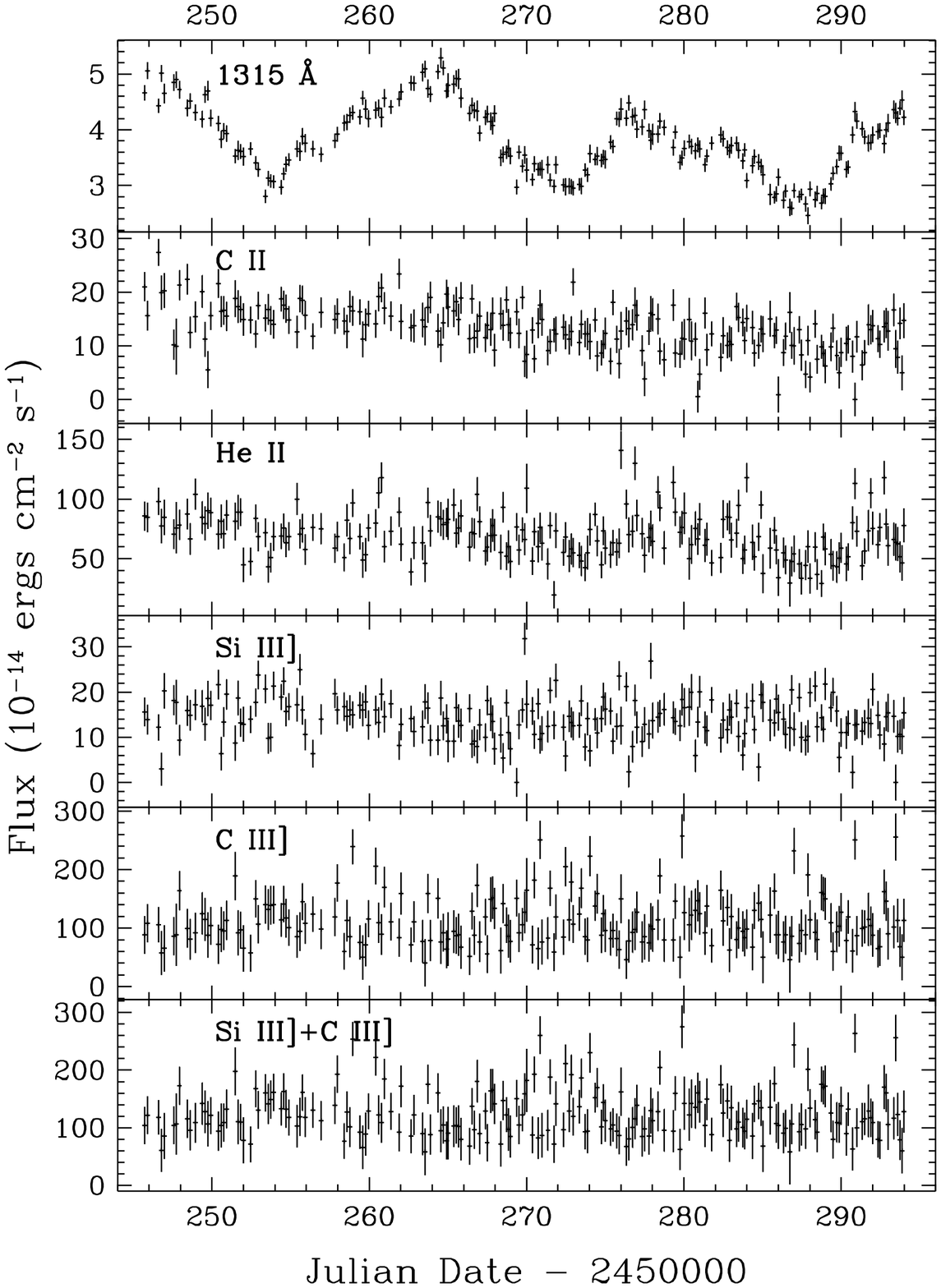"}{6.40 in} {0}{70}{70}{-210}{-47}
\caption{
Light curves for intermediate-strength emission lines from the 1996 campaign
on NGC~7469 extracted from the IUE spectra using the FOS spectrum as a template.
For comparison, the top panel gives the F(1315\AA) continuum light curve
(in units of $10^{-14}~\rm ergs~cm^{-2}~s^{-1}~\AA^{-1}$).
\label{fig9.ps}
}
\end{figure*}

\begin{figure*}[t]
\plotfiddle{"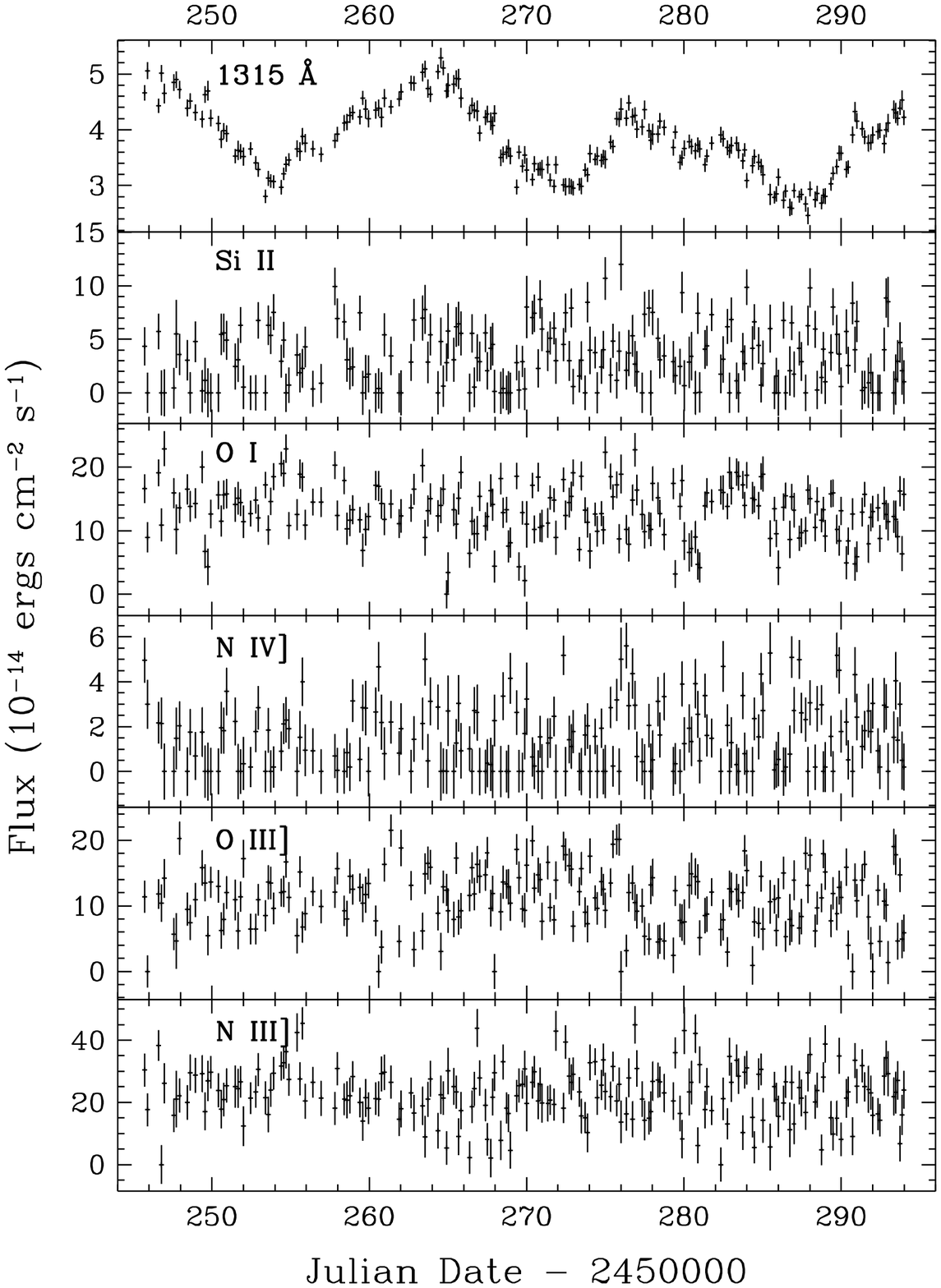"}{6.40 in} {0}{70}{70}{-210}{-47}
\caption{
Light curves for weak emission lines from the 1996 campaign
on NGC~7469 extracted from the IUE spectra using the FOS spectrum as a template.
For comparison, the top panel gives the F(1315\AA) continuum light curve
(in units of $10^{-14}~\rm ergs~cm^{-2}~s^{-1}~\AA^{-1}$).
\label{fig10.ps}
}
\end{figure*}

\subsection{Variability Characteristics}

To quantify the characteristics of the variability in our new measurements,
we use the standard parameters adopted by the International AGN Watch.
We summarize these for all our measured fluxes in Table~7.
The mean flux, $\overline{F}$, and the sample standard deviation (or
root-mean-square flux), $\sigma_F$, have their usual statistical definitions.
The third parameter, $F_{var}$, is the fractional variation in the flux
corrected for measurement errors:
\begin{equation}
F_{var} = {\sqrt{(\sigma_F^2 - \Delta^2)}\over{\overline{F}}},
\end{equation}
where $\Delta^2$ is the mean square value of the individual measurement errors.
The fourth parameter, $R_{max}$, is the ratio of the maximum flux to the
minimum flux.  Note that both of these latter quantities are not very useful
for weaker line fluxes where the measurement
uncertainty is much larger than any intrinsic variations.

For the continuum measurements listed in Table~7,
our fitted fluxes show fractional variations and ratios of maximum to
minimum flux that are slightly greater than or equal to that seen
in the original data, showing that we have probably eliminated
some small amount of less-variable contamination in our fitting process.
In contrast, the fractional variability in the strong
emission lines has
either stayed the same or decreased.  This is likely due to the broad wings
we have included in our line flux measurements.  As one can see in the
rms spectrum shown in

\begin{center}
{\sc TABLE 7\\
Variability Parameters}
\vskip 4pt
\small
\begin{tabular}{l c c c c c }
\hline
\hline
Feature & $N_{data}$ & $\overline{F}^{\rm a}$ &
$\sigma_F^{\rm a}$ & $F_{var}$ & $R_{max}^{\rm b}$\\
\hline
$F_\lambda$(1315 \AA) & 207  &    3.80  &   0.62   &  0.16 &    2.15 \\
$F_\lambda$(1485 \AA) & 207  &    3.85  &   0.56   &  0.14 &    1.95 \\
$F_\lambda$(1740 \AA) & 207  &    3.52  &   0.45   &  0.12 &    1.82 \\
$F_\lambda$(1825 \AA) & 207  &    3.34  &   0.41   &  0.11 &    1.83 \\
                      &      &          &          &       &         \\
Ly$\alpha$            & 207  &  396.77  &  57.09   &  0.12 &    2.13 \\
Ly$\alpha$+N V        & 207  &  504.18  &  65.50   &  0.12 &    1.95 \\
N V                   & 207  &  107.42  &  32.33   &  0.23 &   (7.83) \\
Si IV                 & 207  &   82.96  &  22.12   &  0.21 &    4.45 \\
C IV                  & 207  &  343.10  &  34.44   &  0.07 &    1.80 \\
He II                 & 207  &   69.33  &  19.42   &  0.20 &   (7.19) \\
C III]                & 207  &  110.42  &  42.51   &  0.22 &   (6.36) \\
                      &      &          &          &       &         \\
Si II                 & 207  &    3.38  &   2.84   &  0.62 &  (857) \\
O I                   & 207  &   13.00  &   4.27   &  0.27 &  (10.9) \\
C II                  & 207  &   13.17  &   4.26   &  0.24 &  (52.4) \\
N IV]                 & 207  &    1.54  &   1.50   &  0.68 &  (467) \\
O III]                & 207  &   10.87  &   4.68   &  0.36 &  (23.0) \\
N III]                & 207  &   22.64  &   8.77   &  0.29 &  (21.7) \\
Si III]               & 207  &   14.01  &   4.73   &  0.23 &  (13.8) \\
Si III]+C III]        & 207  &  124.43  &  42.04   &  0.19 &   (4.74) \\
\hline
\end{tabular}
\vskip 2pt
\parbox{3.5in}{
\small\baselineskip 9pt
\footnotesize
\indent
$\rm ^a$Units are $10^{-14}~\rm ergs~cm^{-2}~s^{-1}~\AA^{-1}$
for continuum fluxes and $10^{-14}~\rm ergs~cm^{-2}~s^{-1}$ for line fluxes.\\
$\rm ^b${Uncertain values enclosed in parentheses are dominated
by noise.}
}
\end{center}
\setcounter{table}{7}

\noindent
Figure~1 of Wanders et al., the most variable portion
of each emission line is the line core.  The contrast of this core is less
in the fits we have done using the FOS spectral template.

\subsection{Cross-correlation Analysis}

To re-examine the question of whether there are genuine time delays between the
continuum variations at different wavelengths, we have performed a
cross-correlation analysis of our newly extracted fluxes.
We have used both the interpolation cross-correlation function (ICCF)
(\cite{GS86}; \cite{GP87}),
and the discrete cross-correlation function (DCF) (\cite{EK88}).
Both algorithms use code as implemented by
White \& Peterson (1994)\markcite{WP94}.
We show the derived cross-correlation functions for the continuum fluxes
and bright emission lines in Fig.~11; the CCFs for the weak
emission lines are shown in Fig.~12.

\begin{figure*}[t]
\plotfiddle{"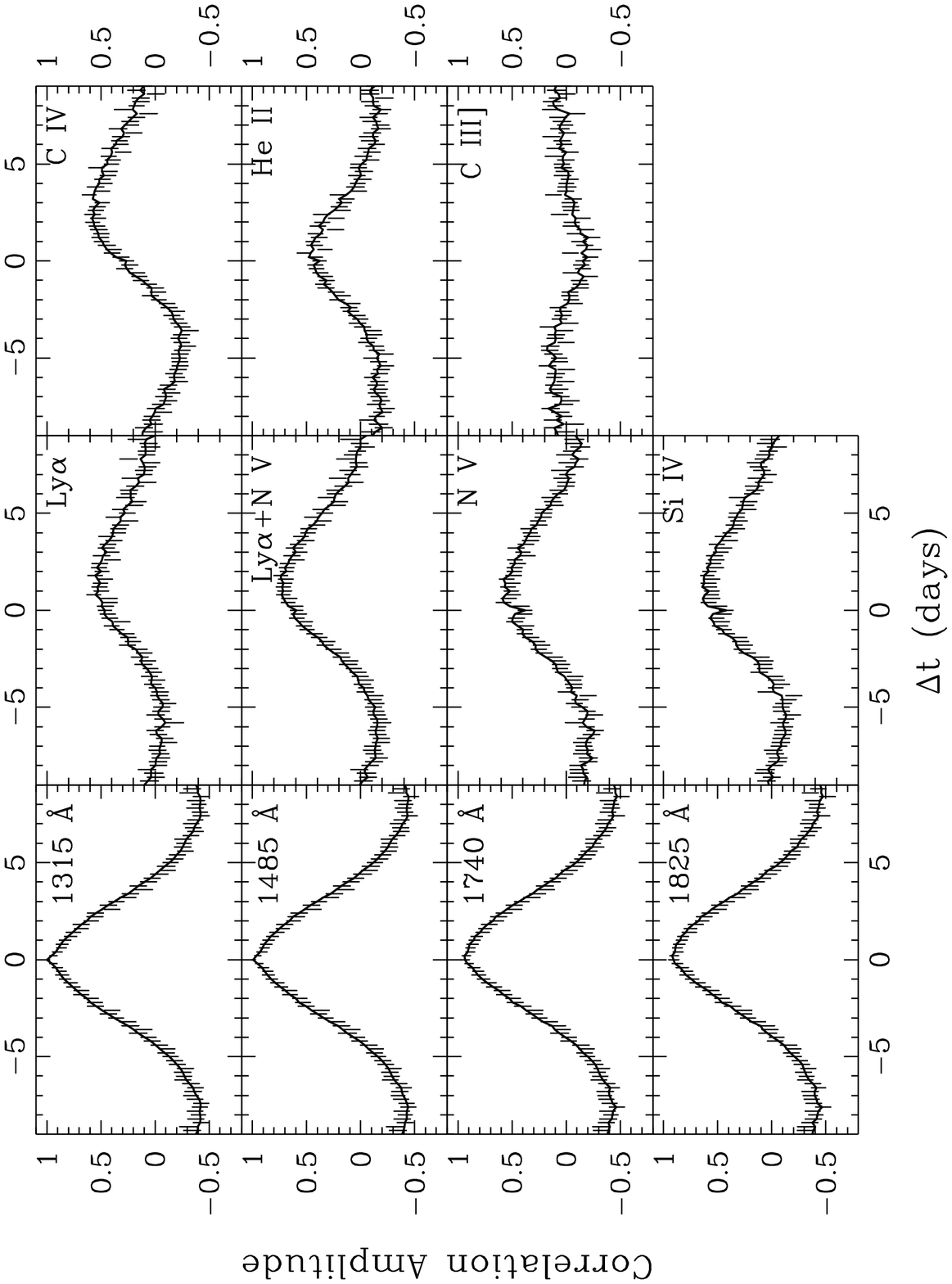"}{3.98 in} {-90}{61}{61}{-250}{338}
\caption{
The continuum and strong emission-line cross-correlation functions for the
fluxes extracted from the IUE spectra of NGC~7469 using the FOS spectrum as
a template.  Each time series has been cross-correlated with the F(1315\AA)
flux series. The interpolated version of the cross-correlation function is
drawn as a solid line; the error bars are the points from the discrete
correlation function.
\label{fig11.ps}
}
\end{figure*}

\begin{figure*}
\plotfiddle{"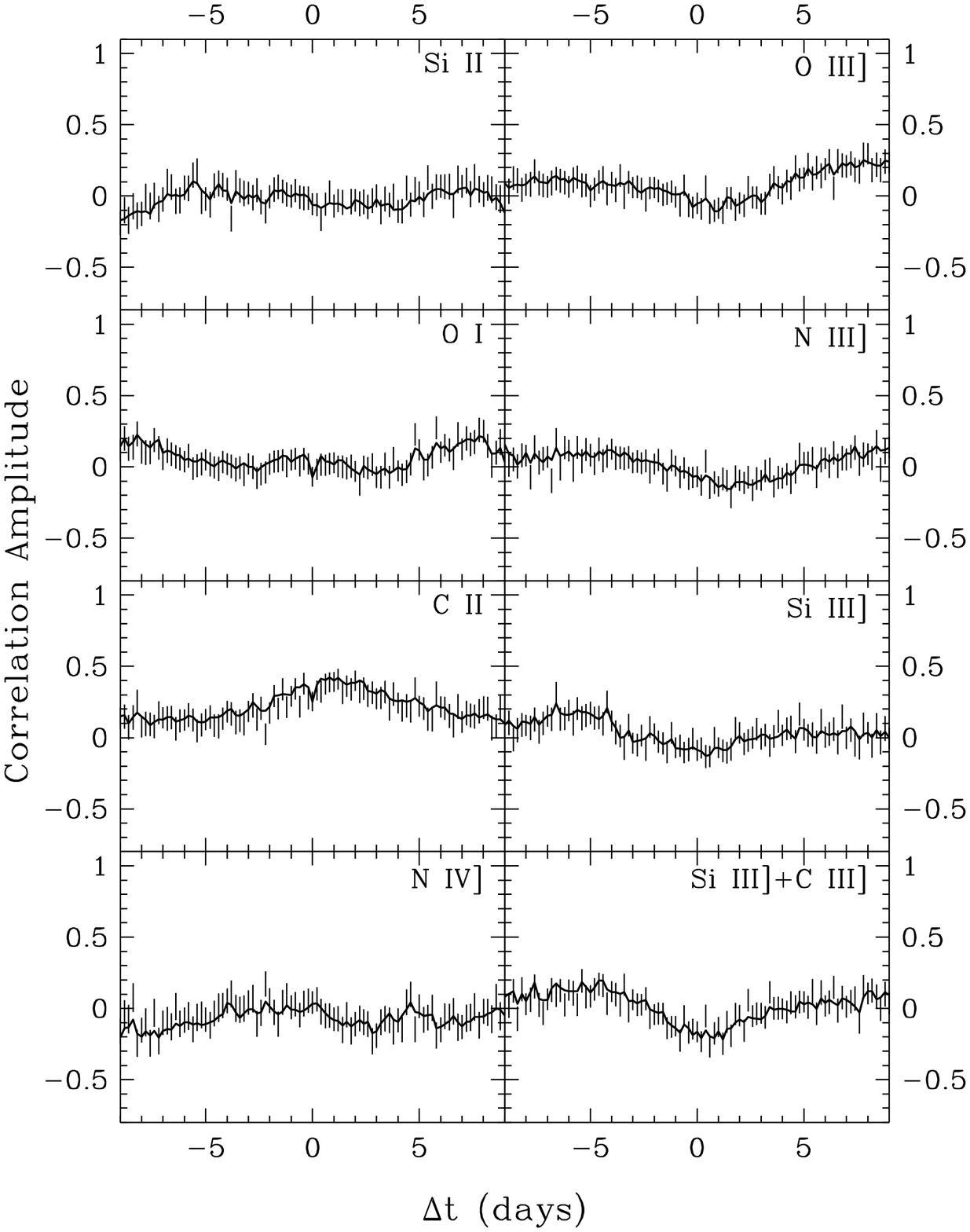"}{5.7 in} {0}{64}{64}{-210}{-51}
\caption{
The weak emission-line cross-correlation functions for the
fluxes extracted from the IUE spectra of NGC~7469 using the FOS spectrum as
a template.  Each time series has been cross-correlated with the F(1315\AA)
flux series. The interpolated version of the cross-correlation function is
drawn as a solid line; the error bars are the points from the discrete
correlation function.
\label{fig12.ps}
}
\end{figure*}

We have made several measurements to quantify the characteristics of
the cross-correlation functions for each measured feature.
In Table~8 we list
the time delay for the centroid of the peak in the CCF, $\tau_{cent}$,
the time delay at which the peak occurs, $\tau_{peak}$,
the peak amplitude, $r_{max}$, of each CCF and
the full-width at half-maximum (FWHM) of the peak.
We calculate the centroids using only CCF values exceeding 80\% of the
peak amplitude.
As the results measured from both the ICCF and DCF curves are nearly identical,
the tabulated numbers are based on the ICCF results.
The error bars for $\tau_{cent}$ and $\tau_{peak}$ are based on
model-independent Monte Carlo simulations
using randomized fluxes and a random subset selection method
as described by Peterson et al. (1998)\markcite{Peterson98}.
Random noise contributions are added to each flux measurement in a light curve,
and a random subset of data pairs is selected for analysis.
This process is repeated many times in a procedure analogous to
``bootstrapping".  Analysis of the resulting distributions from the simulations
leads to the error bars quoted in Table~8 for $\tau_{cent}$
and $\tau_{peak}$.
We note that the smallest of these errors are a factor of $\sim$2 smaller than
the average sampling interval of $\sim0.2$ days, and they are only valid if
there is little variability on timescales shorter than this interval.
High-time-resolution observations of NGC~7469 obtained by Welsh et al.
(1998)\markcite{Welsh98} show that this assumption is valid.

Compared to the results of Wanders et al.\markcite{Wanders97},
the amplitudes of the continuum CCFs are generally slightly higher,
the amplitudes of the emission-line CCFs are generally lower,
and the time delays measured from our CCFs are a bit shorter.
The difference in amplitudes reflects our previous results on the
difference in variability amplitudes--- the continuum measurements
are indeed cleaner, free of low-variability contaminants, and the
emission-line measurements have a greater contribution from the
low-variability broad wings.

The apparently cleaner continuum measurements now permit a critical
re-examination of the question of time delay as a function of wavelength.
Our measured time delays differ from those of
Wanders et al.\markcite{Wanders97}, {\it but the lag
at long wavelengths relative to short wavelengths is still there,
at roughly the same level.}
Relative to the flux at 1315 \AA, the fluxes at 1485 \AA, 1740 \AA\ and
1825 \AA\ have time delays of 0.09, 0.28, and 0.36
days, respectively, compared to the values of 0.19--0.22, 0.32--0.38 and
0.22--0.35 days found by Wanders et al.\markcite{Wanders97}

As noted in \S 3.2, effectively only two of the four continuum flux
measurements are statistically independent due to the
global power-law fit we have used to describe the continuum.
Therefore, although Table~8 shows a monotonically
increasing time delay with wavelength, the monotonic nature
is largely a consequence of the global constraints we have imposed
on the continuum shape.
To assess the influence such a global constraint imposes on our
measured cross-correlation functions, we have performed another
Monte Carlo experiment using the time series of the measured power-law
normalizations and spectral indices.
Starting with the power-law fit parameters builds in the global linkages
between the four wavelength intervals.
As in the random subset selection method described by
Peterson et al. (1998)\markcite{Peterson98},
we chose a random subset from the series of flux normalization points
and indices, preserving the time order of the points.
At the selected time points in a given subset, the normalizations and indices
were assigned random values from a Gaussian distribution with a mean
of the measured value at that time and a dispersion of the $1 \sigma$ error bar.
From these simulated values of normalization and spectral index, we
generated flux points at 1315 \AA, 1485 \AA, 1740 \AA, and 1825 \AA.
We used the ICCF technique to measure the time delay for the centroid of
the CCF peak in these simulated light curves.
For a total of 700 Monte Carlo realizations, relative to the flux at 1315 \AA,
we find median time delays of $0.09 \pm 0.03$, $0.026 \pm 0.07$,
and $0.32 \pm 0.08$ for the fluxes at 1485 \AA, 1740 \AA, and 1825 \AA,
respectively, where the error bars represent the $1\sigma$ confidence intervals.
Thus, from our fits to the IUE data, we can conclude with confidence
that the flux at longer wavelengths lags the flux at shorter wavelengths,
but we cannot conclude that the lag increases as a function of wavelength.
This requires the use of the optical continuum measurements
as discussed in \S 5.1.

\section{ASCA Observations of NGC 7469}

Guainazzi et al. (1994)\markcite{G94} observed NGC~7469 using ASCA between
1993 November 24 and 1993 November 26 for a total exposure time of $\sim$40 ks.
Their analysis of these data note the Fe K$\alpha$ emission feature and a soft excess,
but find no evidence for a warm absorber.  Subsequent analysis of these
same data, benefiting from improved calibration,
by Reynolds (1997)\markcite{Reynolds97} and
George et al. (1998)\markcite{George98}, however, {\it do} clearly detect
absorption edges of {\sc O~vii} and {\sc O~viii} indicative of ionized
absorbing gas.
Reynolds (1997)\markcite{Reynolds97} finds optical depths in the edges of
$\tau_{O7} = 0.17$ and $\tau_{O8} = 0.03$.

To examine whether the UV absorption noted in our FOS spectrum of NGC~7469
could be interpreted in the context of a combined X-ray and UV absorber
(e.g., \cite{Mathur95}), we have retrieved the ASCA data
discussed by Guainazzi et al. from the High
Energy Astrophysics Science Archive Research Center.
These data have been reprocessed with the ``Revision 1" software and
calibration, and we have used the screened event files produced by this
process.  The acceptable SIS data produced by this filtering includes all
data obtained outside of the South Atlantic Anomaly, above a limb angle of
$10^\circ$ from the dark earth and $20^\circ$ from the bright earth, and
in regions of geomagnetic rigidity exceeding 6 $\rm GeV~c^{-1}$.
In addition, we eliminated all data intervals with anomalously high count rates;
the mean rates were 1.5 $\rm cts~s^{-1}$ and 1.1 $\rm cts~s^{-1}$ in the SIS0
and SIS1 detectors, respectively, and we excluded data with
rates $> 3.0~\rm cts~s^{-1}$.
So that Gaussian statistics were applicable in our spectral analysis,
we grouped the extracted spectra for the SIS0 and SIS1 detectors so that each
energy bin contained a minimum of 25 counts.
To avoid the worst uncertainties in the detector response, we restricted our
spectral fits described below to bins with energies in the range
$0.6~\rm keV < E < 10.0~keV$.

Before fitting these data with our warm absorber models, we first verified that
our methods produced empirical results compatible with previous analyses.
We use v10.0 of the X-ray spectral fitting program XSPEC (\cite{xspec96})
for our fits.
We note that a simple power law with absorption by
cold gas gives an unacceptable fit: $\chi^2 = 661.4$ for 424 data bins
and 3 free parameters. Adding a narrow (unresolved) Fe K$\alpha$ line
markedly improves the fit: $\chi^2 = 636.4$ for 424 points and 5 free
parameters. A broad Fe K$\alpha$ line gives further significant improvements,
with $\chi^2 = 565.3$ for 424 points and 8 parameters.
Our best empirical fit to the data is for a power law continuum, absorption
by cold Galactic gas, broad and narrow Fe K$\alpha$ emission from the source,
and two absorption edges representing intrinsic ionized oxygen absorption.
This best fit has $\chi^2 = 482.1$ for 424 points and 12 free parameters.
The best-fit values for the free parameters are summarized in
Table~9.

Our model differs from Reynolds (1997)\markcite{Reynolds97} in that we have
omitted any intrinsic cold-gas absorption, added separate narrow and broad
Fe K$\alpha$ emission lines, permitted the edge absorption energies to vary
freely, and binned our data differently, but our results are comparable.
Our spectral index of $2.14 \pm 0.04$ agrees well with his value of 2.11, and
our edge depth of $0.21 \pm 0.03$ for {\sc O~vii} is in good agreement with
his value of 0.17.  Our optical depth for the {\sc O~viii} edge of
$0.13 \pm 0.03$, however, is larger than Reynold's value of 0.03.
The main reason for this difference is that we have let the edge energies
vary freely, while Reynold's fixed them at their redshifted vacuum energies.

We now describe photoionization models for the ionized absorbing gas that can
be used to evaluate whether the same absorbing medium is responsible for both
the X-ray and the UV absorption.
These models are constructed in the same way as those discussed by Krolik
\& Kriss (1995)\markcite{KK95} and Kriss et al. (1996b)\markcite{Kriss96b}.
For our ionizing spectrum, we use a spectral shape for NGC~7469 based on the
UV and X-ray data discussed here, and the RXTE data presented by Nandra et al.
(1998)\markcite{Nandra98}.
The fit to the mean of the IUE data has a spectral index
(for $F_{\nu} \propto \nu^{-\alpha}$) of 1.087.
The RXTE data has a mean 2--10 keV flux of
$3.4 \times 10^{-11}~\rm erg~cm^{-2}~s^{-1}$.
We re-normalize the ASCA spectrum above (which has $F(2-10) = 3.5 \times
10^{-11}~\rm erg~cm^{-2}~s^{-1}$) to this value.
Since $\alpha_{ox}$ (the effective spectral index between 2500 \AA\ and 2 keV)
is 1.34, we note that the UV and X-ray spectra when extrapolated do not meet
at any intermediate energy--- the ionizing spectrum must steepen between the
UV and X-ray bandpasses. Although the lack of simultaneity between the
UV and X-ray observations may play some role in this mismatch,
this is a common feature of AGN spectra, and
composite QSO spectra suggest that the break occurs around the Lyman limit
(Zheng et al. 1997).  We therefore extrapolate the UV spectrum to the Lyman
limit, and then introduce a spectral break to an index of 1.40 which we
follow to an energy of 0.5 keV, where we then flatten to the X-ray energy index
of 1.14.  Since this spectrum does not diverge to higher energy, we simply
extrapolate this to 500 keV for our photoionization calculations.

As in Kriss et al. (1996b)\markcite{Kriss96b}, we compute our models in
thermal equilibrium, assume constant-density clouds ($n_H = 10^9~\rm cm^{-3}$),
and use the ionization parameter $U = n_{ion} / n_H$, where $n_{ion}$ is the
number density of ionizing photons between 13.6 eV and 13.6 keV illuminating
the cloud and $n_H$ is the density of hydrogen atoms.
We assume that the absorbing medium covers 25\% of the solid angle
around the source.
The transmission of each model is computed so that resonant line scattering
and electron scattering are fully accounted for (\cite{KK95}).
In computing the widths of the resonance lines, we assume that all ions
have turbulent velocities equal to the sound speed in the medium.
The transmission is fully described by two parameters, the total column
density $N_{tot}$, and the ionization parameter $U$.

To fit these models to the ASCA spectra, we assemble
our grid of models into a FITS table to be read into XSPEC,
and we replace the photoionization edges in our empirical model
with the total column density, ionization parameter and redshift
of our warm absorber model grid.
This gives a result comparable in quality to our best empirical fit:
$\chi^2 = 484.2$ for 424 points and 11 free parameters.
Best-fit values for the parameters are given in Table~10,
and the best-fit spectra are illustrated in Fig.~13.

\section{Discussion}

\subsection{Time Delays and the Case for Continuum Reprocessing}

Our newly extracted continuum fluxes for the IUE observations of NGC~7469
in 1996 strengthen the arguments for wavelength-dependent time delays
in the continuum flux from this active galaxy.
In tests performed on the original data set, Wanders et al.
(1997)\markcite{Wanders97} found that contamination by 10\% of
a continuum flux interval by a spectral component with a 2-day lag
could produce a time delay of $\sim$0.2 days in the lag measured
for the continuum flux.
As we note in \S3.1, our model of the FOS spectrum indicates that contamination
by 15--22\%
by weak lines and line wings could be present in the continuum
fluxes for the bands centered at 1315 \AA, 1740 \AA\ and 1825 \AA,
thus implying that the originally measured lags could be 
affected by these other spectral features.
Our new measurements of the IUE spectra greatly ameliorate the potential
level of contamination in the continuum flux points.
The new time delays
we measure are slightly lower (perhaps reflecting
some previous contamination), but the delays are still present,
and they increase with increasing wavelength.

As discussed by Collier et al. (1998)\markcite{Collier98}, a simple
model for radiative reprocessing by a steady-state accretion disk with a
radial temperature profile determined by viscous heat dissipation predicts
that the time delay between different continuum bands should depend on
wavelength as $\tau \propto \lambda^{4/3}$, reflecting the $T \propto R^{-3/4}$
temperature profile and the $\tau = R / c$ dependence of the time delay.
Fig.~14 shows the measured time delay of the UV and optical
continuum points compared to a $\tau \propto \lambda^{4/3}$ curve.
Our new measurements are more consistent with this dependence.

While the UV and optical continuum time delays seem indicative of
radiative reprocessing, the puzzle remains--- what
radiation is being reprocessed?
As Nandra et al. (1998)\markcite{Nandra98} show,
producing the UV and
optical continuum in NGC~7469 via reprocessing of the X-ray radiation
is not energetically feasible, nor does it have the requisite time dependence.
Simultaneous EUVE, ASCA, and RXTE observations of NGC~5548 by
Chiang et al. (1999)\markcite{Chiang99} show that the X-ray variations
lag the EUV variations, and that therefore the EUV cannot be produced via
reprocessing of the hard X-rays.

Nandra et al.'s detailed comparison of the X-ray and UV continuum light curves
in the NGC~7469 campaign shows fairly complex behavior.
The main positive correlation is a 4-day lag in which the UV leads the X-ray
continuum.  This is largely due to the peaks in the UV light curve leading
the X-ray peaks.  In contrast, the light-curve minima are nearly simultaneous.
Nandra et al. (1998)\markcite{Nandra98} suggest that the longer timescale
X-ray variability is due to upscattering of UV seed photons from a variety
of sources at different distances that leads to multiple lags.
At high flux levels, the source of the UV seed photons lies at a distance of
$\sim4$ lt days.  In the flux minima, the seed photons arise closer to the
X-ray production region.
The most rapid X-ray variations are due to variations in the particle
distribution of the scattering medium.
In addition, they suggest that some portion of the EUV continuum is
produced by X-ray reprocessing, and that this is what drives the
line radiation.

Such a scenario poses severe problems for the relative geometry of the
continuum production zone and the broad-line cloud region (BLR), however.
It also is at odds with simultaneous EUVE, ASCA, and RXTE observations of
NGC~5548 (\cite{Chiang99}) that show that
X-ray variations lag the EUV variations, and that therefore the EUV
cannot be produced via reprocessing of the hard X-rays.
In NGC~7469, all the broad lines have measured lags $< 4$ lt days.
If the X-ray radiation is produced closest to the black hole,
the scenario proposed by Nandra et al. would imply that the EUV production
zone and the BLR lie between the X-ray and UV production zones.
Another problem is then one of scale--- 4 lt days from a $10^7$ \Msun\ black
hole corresponds to 7000 gravitational radii $( G M / c^2 )$.
This is a factor of more than 100 higher than the radius at which viscous
dissipation in an accretion disk produces UV and optical continuum radiation.
Producing the majority of UV and optical radiation at such large radii
requires a new, highly efficient dissipation mechanism.

\vbox to 4.1in {
\begin{center}
{\sc TABLE 8\\
Cross-correlation Results}
\small
\vskip 4pt
\begin{tabular}{l c c c c }
\hline
\hline
Feature & $\tau_{cent}$ & $\tau_{peak}$ & $r_{max}$ & FWHM \\
        & (days)        & (days)        &           & (days) \\
\hline
$F_\lambda$(1315 \AA) &  $0.00^{+0.09}_{-0.09}$ & $0.00^{+0.05}_{-0.04}$ 
& 1.00 & 5.11\\
$F_\lambda$(1485 \AA) &  $0.09^{+0.11}_{-0.08}$ & $0.02^{+0.04}_{-0.06}$ 
& 0.99 & 5.12 \\
$F_\lambda$(1740 \AA) &  $0.28^{+0.12}_{-0.13}$ & $0.06^{+0.14}_{-0.02}$
& 0.95 & 5.10 \\
$F_\lambda$(1825 \AA) &  $0.36^{+0.11}_{-0.17}$ & $0.08^{+0.20}_{-0.02}$
& 0.93 & 5.12 \\
$F_\lambda$(4945 \AA) &  $1.17^{+0.55}_{-0.33}$ & $1.33^{+0.35}_{-0.58}$
& 0.89 & 5.60 \\
$F_\lambda$(6962 \AA) &  $1.68^{+1.12}_{-0.82}$ & $1.43^{+1.67}_{-0.53}$
& 0.71 & 6.09 \\
           &       &       &       &       \\
Ly$\alpha$ &  $1.30^{+0.61}_{-0.50}$ & $1.76^{+0.39}_{-1.06}$
& 0.56 & 6.61 \\
Ly$\alpha$+N\,{\sc v} &  $1.48^{+0.28}_{-0.38}$ & $1.71^{+0.04}_{-0.96}$
& 0.74 & 6.37 \\
N\,{\sc v} &  $1.24^{+0.41}_{-0.63}$ & $0.58^{+1.07}_{-0.13}$
& 0.59 & 5.99 \\
Si\,{\sc iv}&  $1.50^{+0.37}_{-0.71}$ & $1.24^{+0.56}_{-0.79}$
& 0.65 & 6.30 \\
C\,{\sc iv}&   $2.81^{+0.62}_{-0.87}$ & $2.24^{+1.26}_{-0.39}$
& 0.59 & 6.38 \\
He\,{\sc ii} & $0.67^{+0.27}_{-0.64}$ & $0.31^{+0.64}_{-0.51}$
& 0.48 & 4.38 \\
C\,{\sc iii}]    &  ---  &  ---  & 0.18 & ---  \\
Si\,{\sc ii}]    &  ---  &  ---  & 0.13 & ---  \\
O\,{\sc i}]      &  ---  &  ---  & 0.23 & ---  \\
C\,{\sc ii} & $1.33^{+0.93}_{-1.86}$ & $1.06^{+1.29}_{-1.56}$
& 0.42 & 7.27 \\
N\,{\sc iv}]     &  ---  &  ---  & 0.14 & ---  \\
O\,{\sc iii}]    &  ---  &  ---  & 0.25 & ---  \\
N\,{\sc iii}]    &  ---  &  ---  & 0.19 & ---  \\
Si\,{\sc iii}]   &  ---  &  ---  & 0.24 & ---  \\
Si\,{\sc iii}]+C\,{\sc iii}]  &  ---  &  ---  & 0.20 & ---  \\
\hline
\end{tabular}
\vbox to 24pt{\vfill}
\end{center}
\setcounter{table}{8}
}

\vbox to 2.9in {
\begin{center}
{\sc TABLE 9\\
Empirical Fit to NGC~7469 X-ray Spectrum}
\vskip 4pt
\small
\begin{tabular}{l c}
\hline
\hline
  Parameter & Best-fit Value\\
\hline
Photon index, $\alpha$     & $2.14 \pm 0.04$ \\
Power law normalization, $F_{1 keV}$ & $(1.36 \pm 0.04) \times 10^{-2}$ \\
                          &          $\rm phot~cm^{-2}~s^{-1}~keV^{-1}$ \\
$N_{HI}$                   & $(4.4 \pm 1.0) \times 10^{20}~\rm cm^{-2}$ \\
Edge Energy, $E_{O7}$\tablenotemark{a}      & $0.685 \pm 0.021$ keV \\
Optical Depth, $\tau_{O7}$ & $0.21  \pm 0.04$      \\
Edge Energy, $E_{O8}$\tablenotemark{a}      & $0.848 \pm 0.021$ keV \\
Optical Depth, $\tau_{O8}$ & $0.13  \pm 0.03$      \\
Narrow Fe Energy\tablenotemark{a}           & $6.345 \pm 0.031$ keV \\
Narrow Fe EW               & $47 \pm 18$ eV \\
Narrow Fe width, $\sigma$  & Fixed at 0.0     \\
Broad Fe Energy\tablenotemark{a}            & $7.03 \pm 0.29$ keV \\
Broad Fe EW                & $3.14 \pm 0.82$ keV \\
Broad Fe width, $\sigma$   & $2.24 \pm 0.48$ keV \\
$\chi^2$/dof               & 482.11/412 \\
\hline
\end{tabular}
\vskip 2pt
\parbox{3.5in}{
\small\baselineskip 9pt
\footnotesize
\indent
$\rm ^a${Energy in the rest frame of NGC 7469, $z=0.0164$.}
}
\end{center}
\vfill
}
\setcounter{table}{9}

\subsection{UV and X-ray Absorption in NGC 7469}

The far-UV spectrum of NGC~7469 as seen with the FOS is typical of other
Seyfert 1s and low redshift AGN.
The intrinsic absorption lines, while hinted at in earlier IUE spectra,
show up clearly.  Like most other Seyfert 1s in which these features are
seen, the equivalent widths (EWs) of 1 \AA\ or less are difficult to detect
in the lower resolution, lower S/N IUE spectra.
With the FOS (and the GHRS), they are now seen to be a common feature
of Seyfert 1s (\cite{Crenshaw99}),
as common as the ``warm absorbers" seen in ROSAT and
ASCA X-ray spectra (e.g., \cite{Turner93}; \cite{Mathur94}; \cite{NP94};
\cite{Fabian94}; \cite{Reynolds97}; \cite{George98}).
Mathur et al. (1994\markcite{Mathur94}, 1995\markcite{Mathur95}) have suggested
a link between the two phenomena, in which the UV absorption lines are
produced by the minority ions in the photoionized gas which is producing
the X-ray absorption.

\vbox to 5.0in {
\vbox to 14pt{\vfill}
\plotfiddle{"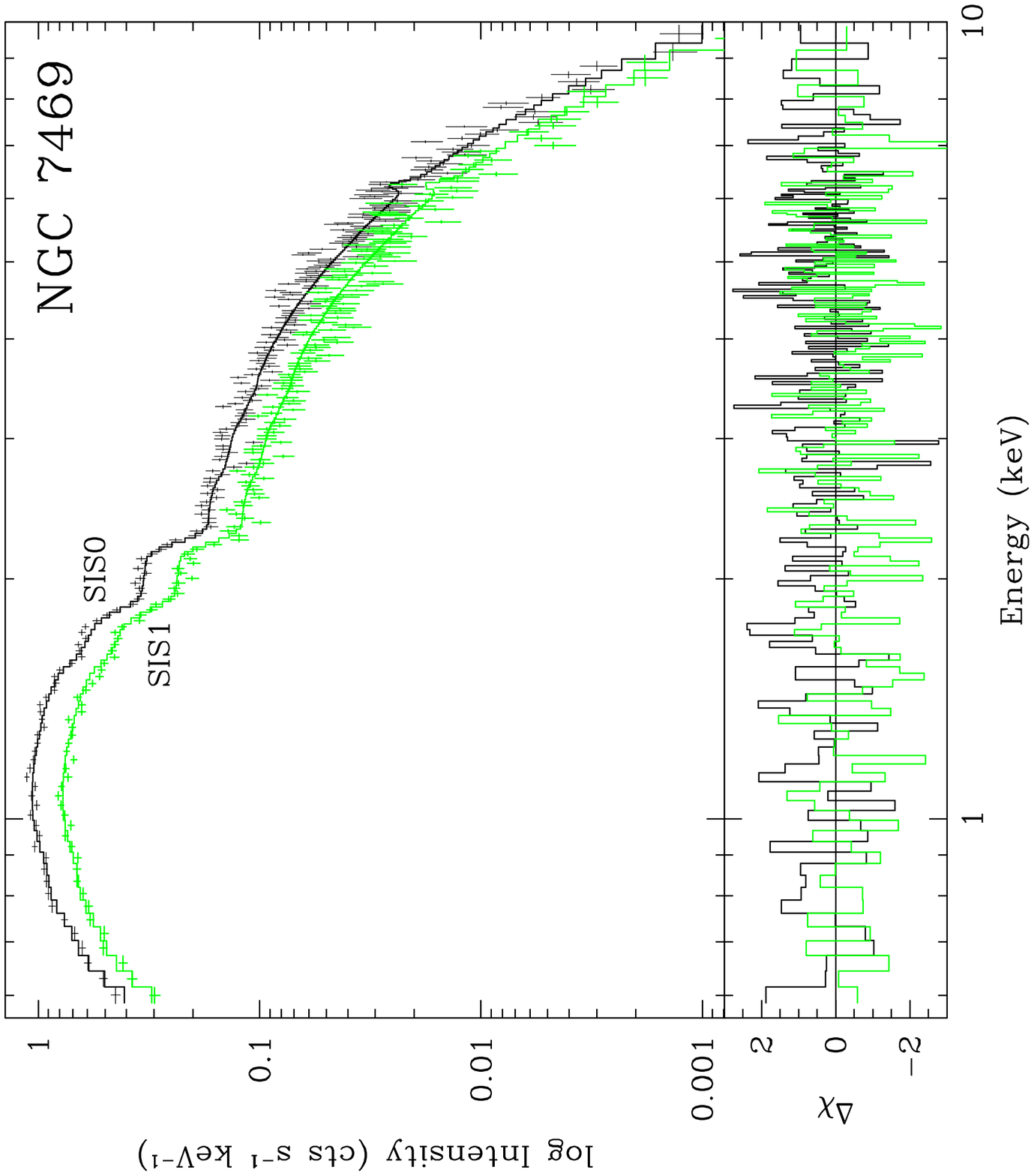"}{3.3 in} {-90}{42}{42}{-165}{265}
\parbox{3.5in}{
\small\baselineskip 9pt
\footnotesize
\indent
{\sc Fig.}~13.---
{\it Upper Panel:} The solid lines are the best-fit warm absorber model for
NGC~7469 folded through the ASCA SIS0 and SIS1 detector responses.
The data points are crosses with 1$\sigma$ error bars.
The model includes a power law with photon index 2.25, absorption by
neutral gas with an equivalent neutral hydrogen column of
$\rm N_H = 5.8 \times 10^{20}~cm^{-2}$,
absorption by ionized gas with a total column density
log $N_{tot} = 21.6~\rm cm^{-2}$
and an ionization parameter of $U = 2.0$,
an unresolved iron K$\alpha$ line at 6.24 keV with an equivalent width of 46 eV,
and a broad (FWHM = 5.9 keV) iron K$\alpha$ line at 6.78 keV with an
equivalent width of 4 keV.
{\it Lower Panel:}
The contributions to $\chi^2$ of each spectral bin are shown.
The solid line is for SIS0 and the dotted line for SIS1.
\label{fig13.ps}
}
\vbox to 14pt{\vfill}
}
\setcounter{figure}{13}

To test whether this is consistent with the strength of the UV absorption lines
seen in NGC~7469, we can calculate the column densities of the UV ion species
using our warm absorber model fit to the X-ray spectrum.
If this single-zone model can simultaneously account for both the X-ray and
the UV absorbers, then the observed EWs of the UV lines should fall on a
single curve of growth consistent with the model.
In Fig.~15 we plot the observed EWs of the Ly$\alpha$, {\sc N~v},
and {\sc C~iv} absorption lines at the column densities predicted by the
best-fit X-ray warm absorber model.
One can see that this is not a self-consistent description of both the
X-ray and UV-absorbing gas.  In particular, the strength of the {\sc C~iv}
absorption lines is much higher than would be predicted for the residual
column in gas ionized sufficiently to produce the observed X-ray absorption.

The observed UV absorption is more consistent with lower column density gas
at a lower ionization parameter.
Fig.~16 shows curves of growth for a photoionization model that
provides the best match to the observed UV absorption lines.
With a total column density of log~$N_{tot} = 19.2~\rm cm^{-2}$ and an
ionization parameter of $U = 0.04$, the observed EWs of Ly$\alpha$, {\sc N~v},
and {\sc C~iv} are nearly all consistent with gas having a Doppler parameter
of $\sim25~\rm km~s^{-1}$.
The total column of this UV-absorbing component is low enough that it would
have negligible effect on the appearance of the X-ray spectrum.
Similarly, as shown by a comparison of Figures 15
and 16, the X-ray absorbing gas makes little contribution to the
UV absorption lines.

\vbox to 2.9in {
\begin{center}
{\sc TABLE 10\\
Warm Absorber Fit to NGC~7469 X-ray Spectrum}
\vskip 4pt
\small
\begin{tabular}{l c}
\hline
\hline
 Parameter & Best-fit Value\\
\hline
Photon index, $\alpha$      & $2.25 \pm 0.06$ \\
Power law normalization, $F_{1 keV}$ & $(1.56 \pm 0.08) \times 10^{-2}$\\
                          &          $\rm phot~cm^{-2}~s^{-1}~keV^{-1}$ \\
$N_H$           & $(5.8 \pm 1.2) \times 10^{20}~\rm cm^{-2}$ \\
Total Column Density, log $N_{tot}$     & $21.6 \pm 0.08$ $\rm cm^{-2}$ \\
Ionization Parameter, $U$             & $2.0  \pm 0.4$      \\
Redshift, $z$             & $0.058  \pm 0.0012$      \\
Narrow Fe Energy          & $6.239 \pm 0.036$ keV \\
Narrow Fe EW              & $46 \pm 34$ eV \\
Narrow Fe width, $\sigma$ & Fixed at 0.0     \\
Broad Fe Energy           & $6.78 \pm 0.33$ keV \\
Broad Fe EW               & $4.1 \pm 1.2$ keV \\
Broad Fe width, $\sigma$  & $2.5 \pm 0.6$ keV \\
$\chi^2$/dof              & 484.20/413 \\
\hline
\end{tabular}
\end{center}
}
\setcounter{table}{10}

\vbox to 4.4in {
\vbox to 14pt{\vfill}
\plotfiddle{"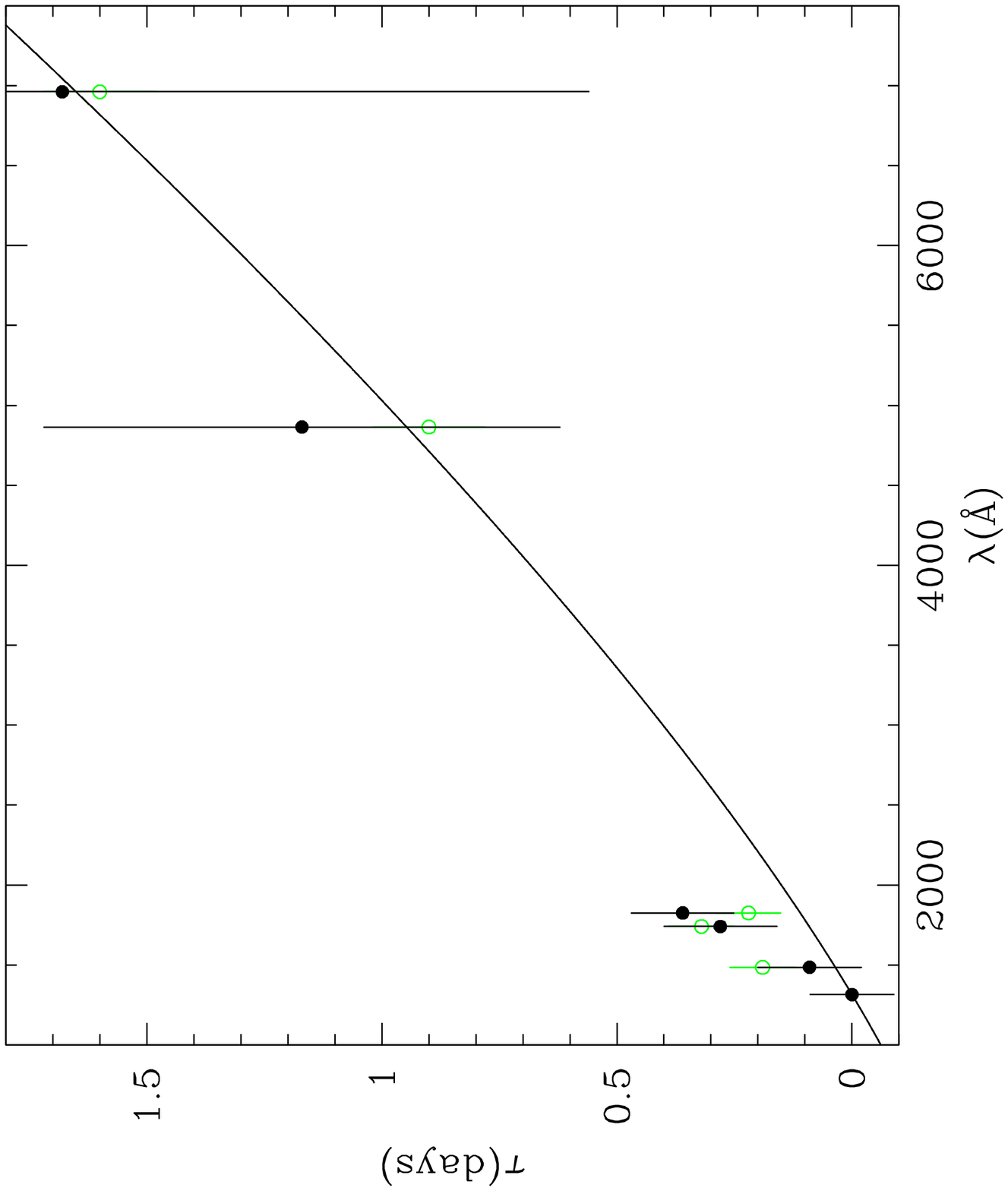"}{3.0 in} {-90}{43}{43}{-170}{260}
\parbox{3.5in}{
\small\baselineskip 9pt
\footnotesize
\indent
{\sc Fig.}~14.---
Time delays vs. wavelength for the IUE continuum bands and the optical
bands presented by Collier et al. (1998).
The gray open circles represent the original measurements of the IUE
data from Wanders et al. (1997) and the optical data from Collier et al. (1998).
The filled black circles are the new measurements from this paper.
The solid line shows the $\lambda^{4/3}$ dependence via the function
$\rm \tau = 3.0 ((\lambda / 10^4 \AA)^{4/3} - (1315 \AA / 10^4 \AA)^{4/3})$.
\label{fig14.ps}
}
\vbox to 14pt{\vfill}
}
\setcounter{figure}{14}

Thus NGC~7469 is yet another instance of a Seyfert galaxy possessing a
complex assortment of absorbing regions.  This was previously shown to be
true for NGC~4151 (\cite{Kriss95}) and for NGC~3516 (\cite{Kriss96a}).
The case of NGC~3516 is particularly illuminating since high resolution
UV spectra show that multiple kinematic components are present, an additional
indication that multiple regions are contributing to the absorption
(\cite{Crenshaw98}).
The Seyferts NGC~4151 (\cite{Weymann97}) and NGC~5548 (\cite{Crenshaw99};
\cite{Mathur99}) also appear kinematically complex when observed at high
spectral resolution.  In fact, in NGC~5548, the prototypical example
of a combined ``XUV" absorber (\cite{Mathur95}),
Mathur et al. (1999)\markcite{Mathur99} now acknowledge that at most one of the
six different kinematic components visible in the high-resolution UV spectrum
actually arises in the X-ray absorbing zone.
Crenshaw and Kraemer (1999)\markcite{CK99} have identified the
kinematic component with the highest blueshift as the one associated
with the X-ray absorber in NGC 5548.

\vbox to 4.8in {
\vbox to 14pt{\vfill}
\plotfiddle{"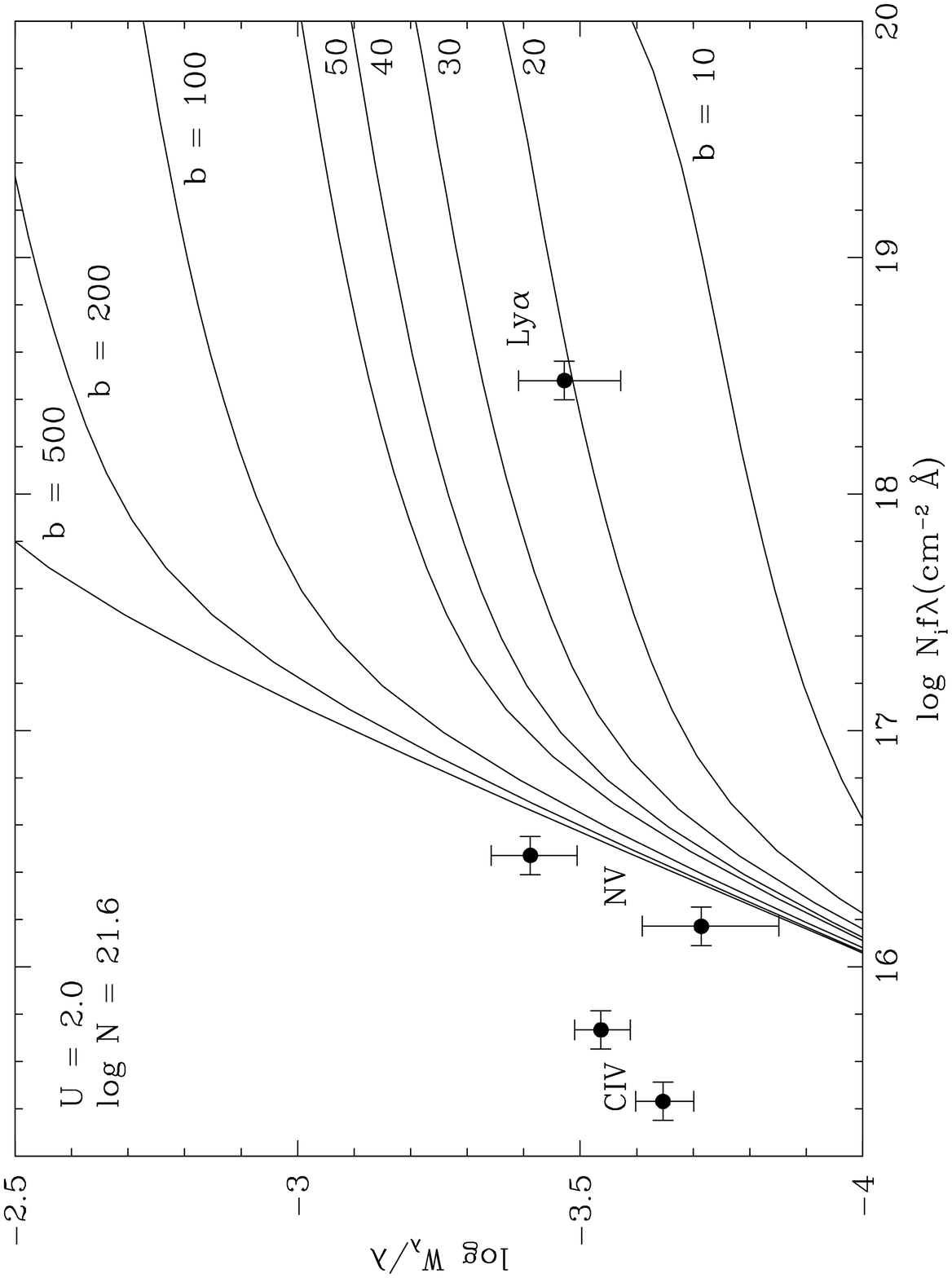"}{2.6 in} {-90}{34}{34}{-130}{215}
\parbox{3.5in}{
\small\baselineskip 9pt
\footnotesize
\indent
{\sc Fig.}~15.---
The observed EWs of the UV absorption lines in NGC~7469 are
plotted on curves of growth using column densities predicted by the single warm
absorber fit to the ASCA X-ray spectrum.
This model has $U = 2.0$ and a total column density of
log $N_{tot} = 21.6~\rm cm^{-2}$.
Points are plotted at a horizontal position determined by the column density
for the given ion in the model with a vertical coordinate determined by the
observed EW for the corresponding absorption line.
The vertical error bars are from Table 2, and the horizontal error bars are
the range in column density allowed by the uncertainty in the fit to the
ASCA spectrum.
The thin solid lines show predicted EWs as a function of column density
for Voigt profiles with
Doppler parameters of $b = 10$, 20, 30, 40, 50, 100, 200,
and $500~\rm km~s^{-1}$.
A model that fits the data would have all points lying on one of these curves.
This model cannot simultaneously match both the UV and the X-ray absorption.
\label{fig15.ps}
}
\vbox to 14pt{\vfill}
}
\setcounter{figure}{15}

\vbox to 3.6in {
\vbox to 14pt{\vfill}
\plotfiddle{"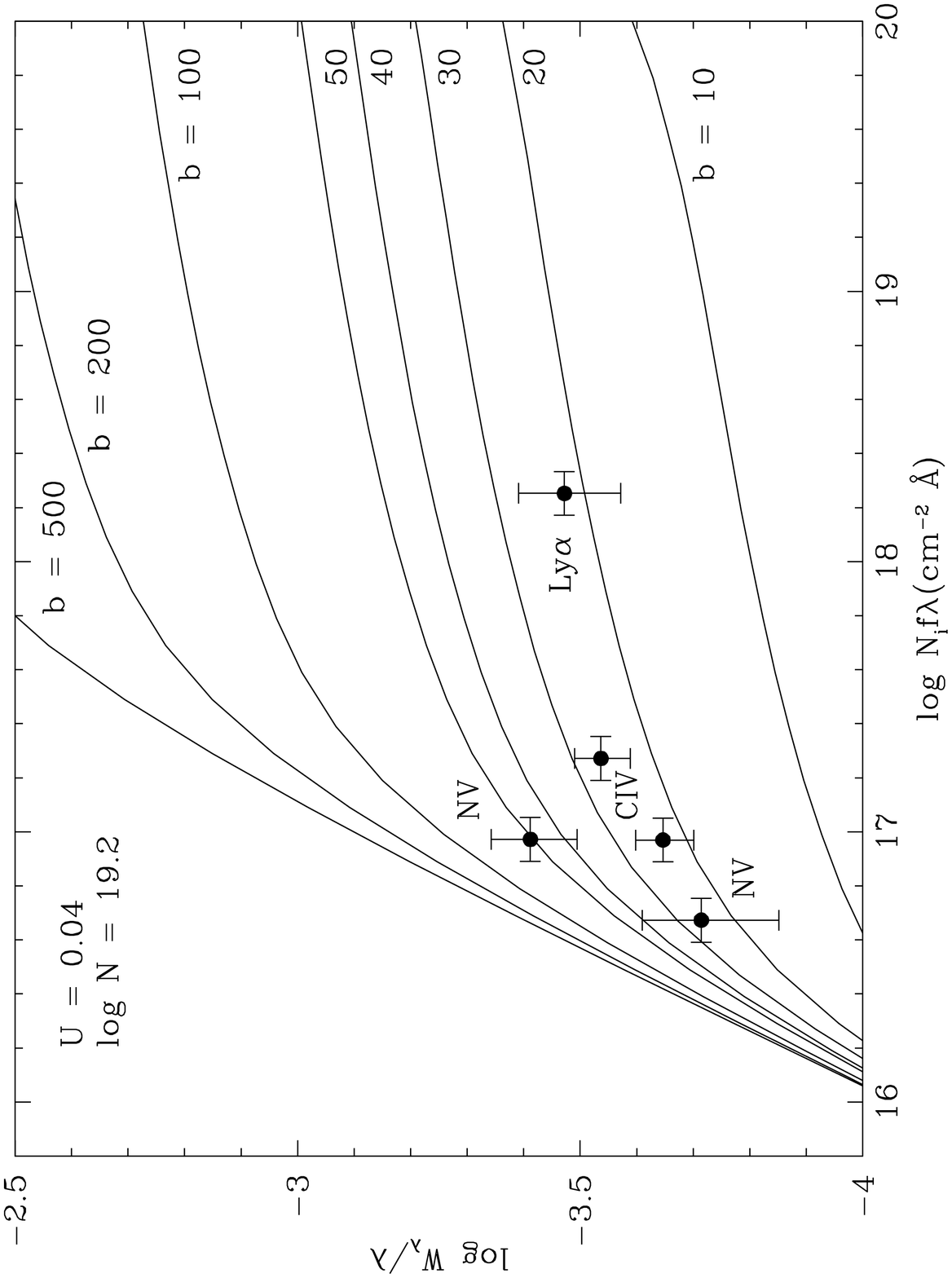"}{2.6 in} {-90}{34}{34}{-130}{215}
\parbox{3.5in}{
\small\baselineskip 9pt
\footnotesize
\indent
{\sc Fig.}~16.---
As in Fig. 15, but for a photoionization model with $U = 0.04$ and
log $N_{tot} = 19.2~\rm cm^{-2}$.
A curve of growth with $b \sim 25~\rm km~s^{-1}$ can match nearly all the
observed EWs.
\label{fig16.ps}
}
\vbox to 14pt{\vfill}
}
\setcounter{figure}{16}

While most Seyferts with UV and X-ray absorption appear to have a complex
assortment of absorbing regions with a broad range of physical conditions,
this does not mean that these physically distinct regions are unrelated.
Since UV and X-ray absorption (or the lack of both) appears to be linked
in most Seyferts (\cite{Crenshaw99}), it is likely that a common mechanism
is responsible for both.
Possibilities for this mechanism include outflows of material ablated from
the obscuring torus (\cite{Weymann91}; \cite{Kriss95}),
or a wind from the accretion disk (\cite{Konigl94}; \cite{Murray95}).
A natural origin for the separate UV and X-ray absorbing clouds
would be to have higher density clumps embedded in a more tenuous wind.
The smaller, higher density clumps would have lower total column densities
and lower ionization parameters, a requirement for the UV absorbers.
The tenuous surrounding wind (which may well have a range of physical
conditions itself, e.g. \cite{Kriss96b}) could be the source for the X-ray
warm absorber.  In such a scenario one might also expect to see correlated
variability in the total column density of the X-ray and UV absorbers
related to ``events" in which new material was ablated from the torus
or accretion disk into the outflowing wind.

\section{Summary}

We have used a high S/N FOS spectrum of NGC~7469 to produce a model template
for extracting deblended emission line and continuum fluxes from the series of IUE
spectra obtained in the 1996 monitoring campaign.
The FOS spectrum shows that ``continuum" windows at 1315 \AA, 1740 \AA\ and
1825 \AA\ used by Wanders et al. (1997)\markcite{Wanders97} in the original
analysis have significant contaminating contributions from the wings of the broad
emission lines and other low-level features such as {\sc O~i} $\lambda 1304$ and
Fe~{\sc ii} emission lines.  Our new extractions for the most part eliminate
these contaminating components from the measured fluxes.
Using these cleaner data, we still find a time delay in the the response of
the continuum flux at longer wavelengths relative to shorter wavelengths.
We find time delays of 0.09, 0.28, and 0.36 days
for the fluxes at 1485 \AA, 1740 \AA\ and 1825 \AA, respectively, relative to
F(1315 \AA).
When combined with the delays measured for the optical continuum by
Collier et al. (1998)\markcite{Collier98}, we find that the wavelength
dependence of the time-delay follows a $\lambda^{4/3}$ relation that
is consistent with the simplest models of radiative reprocessing.

The FOS spectrum of NGC~7469 reveals associated absorption in the high-ionization
lines {\sc N~v}, {\sc C~iv} and Ly$\alpha$, a common feature of Seyfert galaxies
(\cite{Crenshaw99}).  The X-ray spectrum of NGC~7469 also shows evidence for
an ionized absorber (\cite{Reynolds97}; \cite{George98}), and we have analyzed
the UV and X-ray absorbers in the context of a single UV/X-ray absorber
(\cite{Mathur95}).  We find, however, that such a unified description is untenable.
The predicted column densities of UV-absorbing ions in the best-fitting warm
absorber model for the X-ray spectrum imply line strengths well below those
observed.  The UV absorption requires gas with a lower ionization parameter
and lower column density.
Even though the X-ray and UV absorption in this Seyfert and in many others
requires a complex assortment of kinematic components with different physical
conditions, the fact that associated UV absorption and X-ray warm absorbers
are often found in the same objects (\cite{Crenshaw99}) suggests that the material
for each of these absorbers has a common origin.

\acknowledgments
Support for this work was provided by NASA through
grant number GO-06747.01-95A from the Space Telescope Science Institute,
which is operated by the Association of Universities for Research in Astronomy,
Inc., under NASA contract NAS5-26555.
G. Kriss and B. Peterson acknowledge additional support from NASA Long Term
Space Astrophysics grants NAGW-4443 to the Johns Hopkins University and
NAG5-8397 to the Ohio State University, respectively.

\end{document}